\magnification=1200
\pageno=0\footline={\hfill}
\overfullrule=0pt
\hfill UB-ECM-PF-94/19
\bigskip\bigskip
\centerline{\bf Heavy Quark Hadronic Lagrangian}
\centerline{\bf for S-Wave Quarkonium}
 \vskip 1.5cm
\centerline{A. Pineda}
\vskip 0.2cm
\centerline{and}
\vskip 0.2cm
\centerline{J. Soto}
\vskip 0.5cm
\centerline{Departament d'Estructura i Constituents de la Mat\`eria}
\centerline{and}
\centerline{Institut de F\'\i sica d'Altes Energies}
\centerline{Universitat de Barcelona}
\centerline{Diagonal 647, 08028 Barcelona, Catalonia, Spain }
\vskip 1cm

\centerline{\bf Abstract}
\bigskip   \bigskip

\baselineskip=20pt

We use Heavy Quark Effective Theory (HQET) techniques to parametrize
certain non-perturbative effects related to quantum
fluctuations that put both heavy quark and antiquark in quarkonium
almost on shell.
 The large off-shell momentum contributions are calculated using
Coulomb type states. The almost on-shell momentum contributions are
evaluated using an effective 'chiral' lagrangian which incorporates
the relevant symmetries of the HQET for quarks
and antiquarks.
The cut-off dependence of both contributions matches perfectly.
The decay constants and the matrix elements of bilinear currents at zero
recoil are calculated. The new non-perturbative contributions from the
on-shell region are parametrized by a single constant. They turn out
to be $O(\alpha^2/\Lambda_{QCD} a_{n})$, $a_{n}$ being the
Bohr radius and $\alpha$ the strong coupling constant, times
the
non-perturbative contribution coming from the multipole expansion (gluon
condensate). We discuss the physical applications to $\Upsilon$,
$J/\Psi$ and $B_{c}$ systems.

\eject
\footline={\hss\folio\hss}
\baselineskip=20pt

\def\a{\alpha}  \def\b{\beta}  
  \def\e{\epsilon}

\def \Dsl{\,\raise.15ex\hbox{$/$}\mkern-14.5mu {D}}

\def \dsl {\raise.15ex\hbox{$/$}\kern-.57em\hbox{$\partial$}}

\def \Del{\,\raise.15ex\hbox{$/$}\mkern-14.5mu {\Delta}}

\def \v {\rlap\slash{v}}

\def \el {\rlap\slash{\rm e}}

\def \kp {k^{\prime}}

{\bf 1. Introduction}
\bigskip

The so-called Heavy Quark Effective Theory (HQET) [1-5] has become a
standard
tool to study the properties of hadrons containing a single heavy
quark (see [6] for reviews). The hadron momentum is essentially the
momentum of the
heavy quark which may then be considered almost on-shell. The dynamics
becomes independent of the spin and the mass of the heavy quark
giving rise to the so-called
Isgur-Wise symmetries [1,2]. The relevant
modes are momentum fluctuations of the order of $\Lambda_{QCD}
$ which are described by the HQET [3-5]. One cannot actually carry out
reliable perturbative calculations at that scale, but one can certainly
use the Isgur-Wise symmetries to obtain relations between physical
observables.

       \medskip

For hadrons containing two heavy quarks or more the HQET
is not believed to be a suitable approximation. The reason being that
a system of two heavy quarks is mainly governed by
the perturbative Coulomb-type interaction. The relevant modes are
momentum fluctuations of the order of the invers Bohr radius, which is
flavor dependent, and not of the order of $\Lambda_{QCD}$.
 Still, if one is
interested in subleading non-perturbative contributions related to the
"on-shellness" of  the heavy quarks, the HQET may provide some
useful information. Irrespectively of the above, the HQET has already
been used in phenomenological approaches to two heavy quark systems [7].

\medskip

 We shall argue that the leading non-perturbative
contributions in the on-shell region to the quarkonium decay constants
and to the
matrix elements of heavy-heavy currents between quarkonia states can be
described by
a suitably modified HQET. The well-known non-perturbative contributions
in the off-shell region arising from the multipole expansion [8,9] are
$O(\Lambda_{QCD}a_{n}/\a^2)$, $a_{n}$ being the Bohr radius and
$\a$ the strong coupling constant, times the contributions we
find.
The key observation is that when the heavy quarks are almost on-shell
the
non-perturbative effects must be important. In that regime
the multipole expansion breaks down, but it is precisely there where
HQET techniques become applicable.

\medskip

In ref. [10] it was pointed out that when fields describing both heavy
quarks and heavy antiquarks with the same velocity are included in the
HQET lagrangian, the latter has extra symmetries beyond the well known
flavor and spin symmetries [1,2]. In ref. [11] the extra symmetries were
thoroughly analysed (see [12] for related elaborations). It was shown
that they are
spontaneously broken down to the spin and flavor symmetries, even if the
gluons are switched off. The Goldstone
modes turn out to be two particle states with the quantum numbers
of s-wave quarkonia. Translating these findings
 into phenomenologically useful statements was the original motivation
of this work.

\medskip

 The main hypothesis in what follows is that whenever we have a heavy
quark field we may split it in two momentum regimes. The momentum
regime where the heavy quark is almost on shell (small relative three
momentum), and the momentum regime where the heavy quark is off shell
(large relative three momentum).
 The main observation is that the
HQET should always be a
good approximation for a heavy quark in the almost on-shell momentum
regime of QCD [10,12],
 no matter whether the heavy quark is accompained by another heavy
quark in the hadron or not. What makes a hadron containing
a single heavy quark qualitatively different from a hadron containing,
say, two heavy quarks are the large off-shell momentum effects.
 In the former
the large off-shell momentum effects are small
and can be evaluated order by order in QCD perturbation theory
[1,5,13,14].
In
the latter the large off-shell momentum effects are dominant giving rise
to Coulomb-type bound states. However, once this is taken into account
there is
no {\it a priori} reason not to use HQET in the almost on-shell momentum
regime for systems with two heavy quarks. Then the extra symmetries
found in [10,11], which naturally involve quarkonium systems, should be
relevant.

 \medskip

Suppose we have two quarks $Q$ and $Q^{\prime}$ which
are sufficiently heavy so that the
formalism below can be readily applicable. Let us denote by
$\psi_{Q}$,
$\eta_{Q}$, $Q_{Q^{\prime}}^{\ast}$ and $Q_{Q^{\prime}}$
the vector
$\bar Q Q$, pseudoscalar $\bar Q Q$, vector $\bar Q Q^{\prime}$ and
pseudoscalar $\bar Q Q^{\prime}$ states. Our main results follow.

\medskip

(i)The masses do not receive new non-perturbative
contribution from the on-shell momentum region. Consequently,
the leading
non-perturbative correction comes from the multipole
expansion [8,9]. This allows
to extract $m_{Q}$ in a model independent way from
$m_{\psi_{Q}}$, and
hence fix the parameter $\bar \Lambda$ relating $m_{Q}$ with the mass of
the $\bar Q q$ systems [6].

\medskip

(ii) The new non-perturbative effects from the on-shell momentum region
in the decay constants $f_{\psi_{Q}}$,
 $f_{\eta_{Q}}$, $f_{Q_{Q^{\prime}}^{\ast}}$ and $f_{Q_{Q^{\prime}}}$
are given in terms of a single
non-perturbative parameter $f_{H}$.

\medskip

(iii) The new non-perturbative effects from the on-shell momentum
region in the matrix elements of bilinear heavy quark currents
at zero recoil are given in terms of the same non-perturbative
parameter $f_{H}$. In particular, this implies that
 the semileptonic
decays ($m_{Q}>m_{Q^{\prime}}$) $$
\psi_{Q}\, ,\, \eta_{Q}
\longrightarrow
Q_{Q^{\prime}}^{\ast}\, , \,
Q_{Q^{\prime}}
$$
$$
Q_{Q^{\prime}}^{\ast}\, , \,
Q_{Q^{\prime}}
\longrightarrow
\psi_{Q^{\prime}}\, ,\, \eta_{Q^{\prime}}
$$
at zero recoil are known in terms of
$f_{\psi_{Q}}$,
 $f_{\eta_{Q}}$, $f_{Q_{Q^{\prime}}^{\ast}}$ and $f_{Q_{Q^{\prime}}}$.

\medskip

We distribute the paper as follows. In sect. 2 we perform some short
distances calculations in the kinematical region we are interested in.
 In sect. 3 we summarize the main
results of ref. [11] and match the results from sect. 2 with the HQET.
In sect. 4 we construct a hadronic
effective lagrangian for on-shell modes in quarkonium. In sect. 5 we
calculate
the decay constant.
 In sect.
6 we calculate the matrix elements of any bilinear heavy quark current
between quarkonia states. This is relevant for the study of
semileptonic
decays at zero recoil.
 In sect. 7 we briefly discuss the possible use of our formalism for
$\Upsilon$, $B_{c}$, $ B_{c}^{\ast}$, $J/\Psi$ and $\eta_{c}$ physics.
Section 8 is devoted to the conclusions. In Appendix A we show how to
include $1/m$ corrections in the hadronic effective lagrangian for the
on-shell modes. A few technical
details are relegated to Appendix B.

 \bigskip \bigskip

 {\bf 2. Short distance contributions in the on-shell momentum regime}
\bigskip
As mentioned in the introduction, what makes a $\bar Q Q$
system qualitatively
different from a $\bar Q q$ system are the short distance contributions.
In a $\bar Q q $ system these are well understood. They amount to Wilson
coefficients in the currents and in
the operators of the HQET lagrangian, with anomalous dimensions which
are computable in the loop expansion of QCD. For a $\bar Q Q$ system the
short distance contributions cannot be accounted for by just anomalous
dimensions in Wilson coefficients. Indeed, the
anomalous
dimension of a current containing a heavy quark field and a heavy
antiquark field with the same velocity becomes imaginary and infinite
[15]. For large $m_{Q}$, the two quarks
in a $\bar Q Q$ system appear to be very close. Due to assymptotic
freedom the system can be understood in a first approximation as a
Coulomb-type
bound state. In perturbation theory this is
equivalent to sum up
an infinite set of diagrams (ladder approximation) whose kernel is the
tree level one gluon exchange (see [16] for a review).

We shall assume that the dominant short distance contribution to heavy
quarkonia is the existence of Coulomb-type bound states. Typically we
shall be interested in Green functions of the kind
$$  G_{\Gamma}(p_1,p_2):=\int d^4x_1d^4x_2 e^{ip_1x_1+ip_2x_2}
\langle 0\vert T\left\{ {\bar
Q^{a}} \Gamma Q^{b}(0)
{\bar Q}^{b i_1}_{\alpha_1}(x_1)
Q^{a i_2}_{\alpha_2}(x_2)
 \right\} \vert 0 \rangle \,, \eqno (2.1) $$

for the range
of momentum $$ p_1=-m_{b}v-k_1 \quad , \quad p_2=-m_{a}v-k_2 \quad
, \eqno (2.2) $$
$k_1$ and $k_2$ being small.

Since the quarks are very massive, for the range
of momentum (2.2) the leading contribution to (2.1) is only given by the
following ordering
 $$\eqalign{ G_{\Gamma}(p_1,p_2)= & \int d^4x_1d^4x_2
e^{ip_1x_1+ip_2x_2} \theta \left(-max(x_1^0,x_2^0)\right) \cr &
\times \langle 0\vert
\bar
Q^{a} \Gamma Q^{b}(0)T\left\{{\bar Q}^{b i_1}_{\alpha_1}(x_1)
{Q}^{a i_2}_{\alpha_2}(x_2)\right\} \vert 0 \rangle \,.}
\eqno (2.3)$$
We insert the identity between the current and the fields and we
approximate it by the vacuum plus the Coulomb-type states (the
states above threshold
 shall not give contribution when we sit in the relevant pole). We
treat then the fields as being free. $$
1 \simeq \vert 0 \rangle \langle 0 \vert + \sum_{n,s}\int{d^3
\vec P_{n} \over (2\pi )^3 2P_{n}^0}\vert s, \vec P_{n}=m_{ab,n}\vec v
\rangle \langle s,
\vec P_{n}=m_{ab,n}\vec v
 \vert  \eqno (2.4)
$$

The Coulomb state in the center of mass frame (CM) reads
$$\eqalign{ \vert s,
\vec P_{n}=m_{ab,n}\vec v
\rangle                   =&
{1\over \sqrt{N_{c}} }
{m_{ab}^{3\over 2}\over m_{ab,n}}v^0\int {d^3\vec k \over (2\pi)^3}
  \tilde \Psi_{ab,n}(\vec k){1\over \sqrt{2p_1^02p_2^0}} \cr &
\times \sum_{\alpha ,\beta} \bar u^{\alpha} (p_1) \Gamma_{s}
v^{\b}(p_2)
a^{\dagger}_{\a}(p_1)b^{\dagger}_{\b}(p_2) \vert 0\rangle\,,}
\eqno (2.5)
 $$
where
$$ \vec p_1=m_{a}\vec v + \vec k +{\vec k . \vec v \over 1+v^0}\vec v
\quad ,
\quad
\vec p_2=m_{b}\vec v - \vec k -{\vec k . \vec v \over 1+v^0}\vec v\,,
$$
$$  p_1^0=m_{a} v^0 +\vec k . \vec v
\quad ,
\quad
p_2^0=m_{b} v^0 -\vec k . \vec v\,,$$
$$m_{ab}:=m_{a}+m_{b} \quad, m_{ab,n}:=m_{ab}-E_{ab,n} \quad,
\Gamma_{s}=i\gamma_5p_{-} , i\el^{i}p_{-}\,, $$
$$v^2=1\quad, \quad p_{\pm}:={1\pm \v \over 2} \quad, \quad e^{i}.v =
0\,. \eqno(2.6)$$
$E_{ab,n}$, $\Psi_{ab,n}(\vec x)$ and $\tilde \Psi_{ab,n}(\vec k)$ are
the energy, the coordinate space wave function and the momentum space
wave function of a Coulomb-type state with principal quantum number
$n$. $v$ is the bound state 4-vector velocity.
 $a^{\dagger}_{\a}(p_1)$ and $b^{\dagger}_{\b}(p_2)$
are creation operators of particles and anti-particles respectively.
$ u^{\alpha} (p_1)$ and $ v^{\b}(p_2) $ are spinors normalized in such a
way that in the large $m$ limit the following holds
$$ \sum_{\a} u^{\alpha} (p_1) \bar u^{\alpha} (p_1) =p_{+}
\quad , \quad  \sum_{\a} v^{\alpha} (p_1) \bar v^{\alpha} (p_1) =-p_{-}
\,.\eqno (2.7)
$$
 Choosing the momenta as in (2.6) is crucial in
order to take into account that the
CM of the bound state moves with a fix velocity $v$ with respect to the
laboratory frame [17]. (2.5) has the usual relativistic normalization
$$
\langle  s,
\vec P_{n}=m_{ab,n}\vec v
\vert
 r,
\vec P_{m}=m_{ab,m}\vec v^{\prime}
\rangle                   =
2m_{ab,n}v^0 (2\pi)^3\delta^{(3)}(m_{ab,n}(\vec v -\vec v^{\prime}))
 \delta_{nm}\delta_{rs}\,.
\eqno (2.8)
$$
We have to consider the following kind of matrix elements
$$  \eqalign{
&\langle  s,
 m_{ab,n}\vec v
\vert
Q_{\a_2}^{a}(x_2) \bar Q_{\a_1}^{b} (x_1) \vert 0 \rangle
= e^{im_{ab,n}v.X}
\langle  s,
m_{ab,n}\vec v
\vert
Q_{\a_2}^{a}(x_2-X) \bar Q_{\a_1}^{b} (x_1-X) \vert 0 \rangle
\cr & \,=
e^{im_{ab,n}v.X}
{m_{ab}^{3\over 2}\over m_{ab,n}}
 (\bar \Gamma_s )_{\a_2\a_1}
\int {d^3\vec k \over (2\pi)^3}
 \tilde \Psi_{ab,n}^{\ast}(\vec k)
e^{i(\vec k .\vec v x^0 - \vec x (\vec k +{\vec k .\vec v \over
1+v^0}\vec v)) } \,, }
\eqno (2.9)
 $$
$$X={m_{a}x_1+m_{b}x_2 \over m_{ab}} \quad ,\quad x=x_1-x_2 \,.$$
where it is essential to extract the CM dependence in the fields {\it
before} using the explicit expression (2.5) for the calculation of
(2.9).
As mention above the states $
\vert s,
m_{ab,n}\vec v
\rangle
   $ have the explicit expresion (2.5) only
in the CM frame [16,17].
Factors of the kind $m_{ab}/ m_{ab,n}$ appearing in several
expressions above have been approximated to $1$ in the rest of the
paper.
Finally, performing the $x_{1}$, $x_{2}$ integral and taking into
account that
$$
\sum_s
(\Gamma_{s} )_{\alpha_2\alpha_4}
(\bar\Gamma_{s})_{\alpha_1\alpha_3}=-2(p_{+})_{\alpha_2\alpha_3}(p_{-})_{
\alpha_1\alpha_4} \eqno(2.10) $$
we obtain
$$\eqalign{ G_{\Gamma}(p_1,p_2)= & \sum_{n}
\tilde \Psi_{ab,n}^{\ast}(0)
 \Psi_{ab,n}(0) (p_{-}\Gamma p_{+})_{\alpha_2\alpha_1}
\delta_{i_1 i_2} \cr & \times
{1\over v.k_2+{m_{a}\over m_{ab}} E_{ab,n} +i\epsilon}\,
{1\over v.k_1+{m_{b}\over m_{ab}} E_{ab,n} +i\epsilon} \,,}
\eqno (2.11)
 $$
In the last expression we approximated
$\tilde \Psi_{ab,n}(e^{i}. k) \simeq
\tilde \Psi_{ab,n}(0)$ (we neglect $O(({n\mid e^{i}.k \mid\over
m\alpha})^2)$).
In (2.11) there is a sum over an infinite number of poles.
Each term in the sum corresponds
to a Coulomb-type bound state. At the hadronic level we want to describe
only one of those states.
 This is
achieved by tunning the external momenta to sit on the relevant pole.
Suppose we are interested in $\psi_{Q} (n)$ state. Then we take
$$ k_1=
k_1^{\prime}-{m_{b}\over m_{ab}} E_{ab,n} v \quad ,\quad
k_2=
k_2^{\prime}-{m_{a}\over m_{ab}} E_{ab,n} v \,,
\eqno (2.12)
$$
so that in the limit $k_{i}^{\prime}\rightarrow 0$, ($i=1,2$) we obtain
$$\eqalign{  G_{\Gamma}(p_1,p_2)=&
\tilde \Psi_{ab,n}^{\ast}(0)
 \Psi_{ab,n}(0)
 (p_{-}\Gamma p_{+})_{\alpha_2\alpha_1}
\delta_{i_1 i_2}
\cr & \times
{1\over v.\kp_2 +i\epsilon} \,
{1\over v.\kp_1 +i\epsilon} \,. }
\eqno (2.13)
$$

Notice from (2.2) and (2.12) that we must subtract from the momentum of
the quark $(m_a-{m_a \over m_{ab}}E_{ab,n})v$ in order to get an
expression suitable to be reproduced in the HQET. This may be
interpreted as if integrating out off-shell short distance degrees of
freedom produces an effective mass for the almost on-shell modes of a
heavy quark inside quarkonium. This effective mass depends on the
precise bound state the quark is in. We are almost on-shell when
$v.k_{i}^{\prime},e^{j}.k_{i}^{\prime} \sim \Lambda_{QCD}$ (i=1,2).

This restricts the vality of our approximation to the case
$E_{ab,n}\sim \mu_{ab}\alpha^2/n^2\gg\Lambda_{QCD}$ ( $\mu_{ab}$ is
the reduced mass), otherwise momentum fluctuations of the order
of $\Lambda_{QCD}$ would take us from one pole to another. Notice also
that for arbitrary large but fix $\mu_{ab}$ there is always an $n$ where
this
approximation fails. Therefore we shall always be dealing with a finite
number of low laying energy levels.
\medskip
Consider the four-point function.
$$
\eqalign{ G(p_1,p_2,p_3,p_4):=&\int d^4x_1 d^4x_2d^4x_3d^4x_4
e^{ip_1x_1+ip_2x_2+ip_3x_3+ip_4x_4} \cr   & \times
\langle 0 \vert T\left\{
Q^{b i_1}_{\alpha_1}(x_1)
Q^{a i_2}_{\alpha_2}(x_2)
{\bar Q}^{a i_3}_{\alpha_3}(x_3)
{\bar Q}^{b i_4}_{\alpha_4}(x_4) \right\} \vert 0 \rangle \,.} \eqno
(2.14) $$
For the momenta
$$ \eqalign{
& p_1=-(m_{b}-{m_{b}\over m_{ab}}E_{ab,n})v-k^{\prime}_1 \,,
 \cr & p_2=(m_{a}-{m_{a}\over m_{ab}}E_{ab,n})v+k^{\prime}_2 \,,
 \cr & p_3=-(m_{a}-{m_{a}\over m_{ab}}E_{ab,n})v-k^{\prime}_3 \,,
 \cr & p_4=(m_{b}-{m_{b}\over m_{ab}}E_{ab,n})v+k^{\prime}_4  \,,
 }\eqno (2.15)
$$
($k_{i}^{\prime}\rightarrow 0$ , $i=1,...,4$) working in the same way we
obtain $$
\eqalign{
G(p_1,p_2,p_3,p_4)&=(2\pi)^4
\delta^{(4)}(-k_1^{\prime}+k_2^{\prime}-k_3^{\prime}
+k_4^{\prime}) {i\over 2N_{c}}
\sum_{\Gamma_{n}=i\gamma_5p_{-} , i\el^{i}p_{-}}
(\Gamma_{n} )_{\alpha_2\alpha_4}
(\bar\Gamma_{n})_{\alpha_1\alpha_3}
\cr & \times
\delta_{i_1 i_3}
\delta_{i_2 i_4}
\tilde \Psi_{ab,n}^{\ast}(0)
\tilde \Psi_{ab,n}(0)
{1\over v.k_3^{\prime}+i\epsilon}\,
{1\over v.k_1^{\prime}+i\epsilon} \left(
{1\over v.k_2^{\prime}+i\epsilon}  +
{1\over v.k_4^{\prime}+i\epsilon}  \right) \,.}     \eqno (2.16)
$$
We shall see in the next section that (2.13) and (2.16) can be reproduced
(with suitable changes) by a HQET for quarks and antiquarks.

\bigskip
{\bf 3. HQET for quarks and antiquarks}
\bigskip

The lagrangian of the HQET for quarks and antiquarks moving at the same
velocity
$v _ {\mu}$ ($ v _ {\mu} v ^{\mu} =1 $)
 reads [4]
$$
L_v = i \bar h_v \v v _ {\mu} D ^ {\mu} h _ v \,= \, i \bar h ^+ _
v v \cdot D h ^+ _v \,- \, i \bar h^- _v v \cdot D h ^- _ v
\,,
\eqno(3.1) $$
where $ h _ v = h ^+ _ v + h ^- _ v $ and $ h ^{\pm} _ v = { 1 {\pm} \v
\over 2 } h _ v $. $ h ^+ _ v $ contains annihilation operators of
quarks with small momentum about $ m v _ {\mu} $ and $ h ^- _v$ contains
creation operators of anti-quarks again with small momentum about $
m v _ {\mu} $. $D_{\mu}$ is the covariant derivative containing the
gluon field.

The quark and antiquark sector of (3.1) are independently invariant
under the well-known spin and flavour symmetry [1,2,4]
$$
h ^{\pm} _v \rightarrow e ^{i \epsilon ^i _ {\pm} S ^{\pm} _ i }
h ^{\pm} _v \; \; \;{\rm and } \;\;\; \bar h ^{\pm}_{v} \rightarrow \bar
h ^{\pm}_{v} e ^{ -i \e ^i _ {\pm} S ^{\pm} _ i }\; ,
\eqno(3.2)$$
where $ S ^{\pm} _i = i \e _ {ijk} [ \el _ j , \el _k ] ( 1 \pm \v)
/2 \,$, with $ e ^{\mu} _ j \;, j = 1,2,3 $ being an orthonormal set of
space like vectors orthogonal to $ v _\mu \; $, and
$$
h ^{\pm} _v  \rightarrow e ^ {i \theta _{\pm}}  h ^{\pm}
_v  \; \;\; {\rm and } \;\;\; \bar h ^{\pm}_ v \rightarrow \bar h
^{\pm}_v  e ^{-i \theta _ {\pm}  }\; .
\eqno(3.3)$$
 $ \epsilon ^i _ {\pm} $ and $
\theta _ {\pm} $ are arbitrary real numbers corresponding to
the parameters of the transformations.

The lagrangian (3.1) is also invariant under the following set of
transformations
$$
h _v \rightarrow e ^{i \gamma _ 5 \epsilon } h _ v \;\; \;\;{\rm ;}
\;\;\;\;
\bar h _ v \rightarrow \bar h _ v e ^{i \gamma _ 5 \epsilon } \, ,
\eqno(3.4)$$
$$
h _ v \rightarrow e ^{ \gamma _ 5 \v \e } h _v \;\;\;\;{ \rm ;} \;\;
\;\; \bar h _v \rightarrow \bar h _v e ^{ \gamma _ 5 \v \e } \, ,
\eqno(3.5)$$
$$
h _v \rightarrow e ^{ \e ^i \el _ i } h _ v \;\;\;\; {\rm ;}\;\; \;\;
\bar h _ v \rightarrow \bar h _ v e ^{ \e ^i \el _ i } \, ,
\eqno(3.6)$$
$$
h _ v \rightarrow e ^{ i \e ^i \el _ i \v } h _ v \;\; \;\; {\rm ; }
\;\;\;\; \bar h _v \rightarrow \bar h _v e ^{i \e ^i \el _ i \v } \, .
\eqno(3.7)$$
The whole set of transformations (3.2)-(3.7) corresponds to a $U(4)$
symmetry for a single flavour. For $N_{hf}$ heavy flavours they
correspond to a $U(4N_{hf})$ group. In the latter case $h_{v}$ must be
considered a vector in flavour space and the parameters of the
transformations (3.2)-(3.7) as hermitian matrices in that space.
\medskip

 When
the gluons are switched off it
is easy to prove that the $U(4N_{hf})$ symmetry breaks spontaneously
down
to $U(2N_{hf})\otimes U(2N_{hf})$ (see [11]). The following currents
correspond to the broken generators
$$
  j _{5\pm}^{ab}:=\bar h_{v}^{a}i \gamma_5 p_{\pm} h_{v}^{b}  \quad
{\rm and}\quad
 {j _{5\pm}^{ab}}^{i}:=\bar h_{v}^{a}i {\el } _{i} p_{\pm} h_{v}^{b} \,
, \eqno(3.8)$$
$a,b,c...=1,...N_{hf}$ are flavour indices. They transform according to
two
four dimensional irreducible representations of $U(2N_{hf})\otimes
U(2N_{hf})$.
In what follows we are going to assume that the situation above is not
modified when soft gluons are switched on.
The currents (3.8) have the quantum numbers of pseudoescalar and
vector quarkonium respectively. The heavy quark and antiquark fields
interact with soft gluons according to the lagrangian (3.1). For soft
gluons, perturbation theory cannot be realiable applied. However, one
can
use effective lagrangian techniques, which fully exploite the symmetries
above, to parametrize the non-perturbative contributions in this region.
This shall be done in section 4.

 \medskip
For further purposes let us carry out some leading order perturbative
calculations. Consider first
$$\eqalign{ G_{\Gamma \Gamma^{\prime}}(k)=& \int d^4xe^{-ik.x}
\langle 0 \vert T\left\{
{\bar h_{v}^{a-}}\Gamma
{h_{v}^{b+}}(0)
{\bar h_{v}^{b+}}  \Gamma^{\prime}
{h_{v}^{a-}}(x)
\right\}\vert 0 \rangle          \cr
= &-iN_{c}
{\mu^3 \over 6\pi^2}
tr\left(p_{+}\Gamma^{\prime}p_{-}\Gamma \right)
{1\over v.k+i\epsilon } \, ,}
\eqno (3.9)
$$
where $\mu$ is an ultraviolet symmetric cut-off in three-momentum
(see [11] for more details). Consider also
$$              \eqalign{ &
G_{\Gamma \Gamma^{\prime} \Gamma^{\prime\prime}} (\kp_1,\kp_2) \cr &=
\int d^4x_1
d^4x_2 e^{i\kp_1x_1-i\kp_2x_2}
\langle 0\vert T\left\{
\bar h_{v}^{a-}\Gamma^{\prime\prime}
 h_{v}^{b+}(x_1)
\bar h_{v}^{b+}\Gamma h_{v}^{c+}(0)
\bar h_{v}^{c+}\Gamma^{\prime} h_{v}^{a-}(x_2)
\right\} \vert 0 \rangle
  \cr &=
N_{c} {\mu^3\over 6\pi^2}
tr\left(p_{-} \Gamma^{\prime\prime} p_{+} \Gamma p_{+}
\Gamma^{\prime}\right)
{1 \over v.\kp_1
+i\epsilon}\,
{1 \over v.\kp_2
+i\epsilon}\, . }
\eqno (3.10)
$$
The flavor indices ($a$ , $b$ , $c$) are not summed up unless
otherwise indicated. Colour indeces are not explicitly displayed
in the colour singlet currents. Otherwise they will be denoted by $i_1$
, $i_2$, ...$=1 .... N_{c}$,
 $N_{c}$ being the number of colours.
 We
shall drop the subscript $v$ from $h_{v}$ and change the superscripts
$\pm$ into subscripts in the following.

\medskip

The strong cut-off dependence of (3.9)-(3.10) is puzzling. We shall see
later on that it cancels against suitable short distance contributions.
\medskip

As claimed before, it is easy to see that (2.13) is reproduced by the
following
HQET Green function at tree level
$$  G_{\Gamma}(k^{\prime}_1,k^{\prime}_2)=
\int d^4x_1d^4x_2 e^{-ik^{\prime}_1x_1-ik^{\prime}_2x_2}\langle 0
\vert T\left\{ C_{\Gamma}\bar
h^{a} \Gamma h^{b}(0){\bar h}^{b\;i_1}_{+\alpha_1}(x_1)
h^{a\;i_2}_{-\alpha_2}(x_2)\right\} \vert 0 \rangle\eqno (3.11) $$
with $C_{\Gamma}$ being a Wilson coeficient.
$$C_{\Gamma}=
\tilde \Psi_{ab,n}^{\ast}(0)
 \Psi_{ab,n}(0)\,. \eqno (3.12)$$

Analogously, (2.16) is reproduced in the HQET by
\footnote{*}{One may be tempted to include (3.13) as a perturbation
in the HQET lagrangian. This is not quite correct. The Green function
$$
\eqalign{ G(k^{\prime}_1,k^{\prime}_2,\kp_3,\kp_4)=&
\int d^4x_1
d^4x_2d^4x_3d^4x_4 e^{-i\kp_1x_1+i\kp_2x_2-i\kp_3x_3+i\kp_4x_4} \cr   &
\times \langle 0 \vert T\left\{
{h}^{b\;i_1}_{-\alpha_1}(x_1)
{h}^{a\;i_2}_{+\alpha_2}(x_2)
{\bar h}^{a\;i_3}_{+\alpha_3}(x_3)
{\bar h}^{b\;i_4}_{-\alpha_4}(x_4) \right\}
 \vert 0 \rangle }
$$
gives a non-zero contribution in the HQET which does not correspond to
(2.14)-(2.16). It is (3.13) which gives the leading contribution to
(2.14) in the HQET and
hence the last term in (3.13) must not be included in
the lagrangian. This means that unlike
in the case of heavy-light systems, the short distance effects here
cannot always be
accounted for by only modifications of the currents and the lagrangian,
as we may have na\"\i vely expected.
We have to content ourselves by identifying for a given Green function,
the Green function in the HQET which gives the same result.}

$$\eqalign{ G(k^{\prime}_1,k^{\prime}_2,\kp_3,\kp_4) & = \int d^4x_1
d^4x_2d^4x_3d^4x_4 e^{-i\kp_1x_1+i\kp_2x_2-i\kp_3x_3+i\kp_4x_4} \cr   &
\times \langle 0 \vert T\Biggl\{
{h}^{b\;i_1}_{-\alpha_1}(x_1)
{h}^{a\;i_2}_{+\alpha_2}(x_2)
{\bar h}^{a\;i_3}_{+\alpha_3}(x_3)
{\bar h}^{b\;i_4}_{-\alpha_4}(x_4) \cr &
\times i\int d^4y \left(-{1\over 2N_{c}}
\tilde \Psi_{ab,n}^{\ast}(0)
\tilde \Psi_{ab,n}(0) \right)
\sum_{\Gamma_{n}}
\bar h^{a}\Gamma_{n} h^{b}(y)
iv.\partial (
\bar h^{b}\bar \Gamma_{n}h^{a}(y))
 \Biggr\} \vert 0 \rangle \,.}
\eqno (3.13) $$

\bigskip
{\bf 4. Effective hadronic lagrangian for the on-shell contributions of
s-wave quarkonia} \bigskip

We have seen that for the on-shell kinematical regime certain
correlators can be reproduced in the HQET. We shall see in the sect. 5
and 6 that the contributions from this region to the decay constants
and matrix elements reduce to the evaluation of heavy quark-antiquark
currents in the HQET. For the range of momentum we are interested in
these Green functions cannot reliable be evaluated in perturbation
theory. We shall use in this section effective lagrangian techniques,
very similar to those used in Chiral perturbation theory, to parametrize
the nonperturbative contribution.
\medskip
There
are well-known rules [18] (see also [19]) to construct phenomenological
lagrangians for Goldstone bosons  associated to the symmetry
breaking of a group {\sl G} down to a subgroup {\sl H} for relativistic
theories. These rules need two slight modifications to become applicable
to our case:

 (i) The HQET is formally relativistic only after assigning
transformation properties to the fix velocity $v^{\mu}$. We must
take
into account that the velocity $v^{\mu}$ as well as the $e_{\mu}^{i}$
can also be used to build up relativistic invariant terms.

 (ii) The HQET is not only globally $U(4N_{hf})$ invariant, but locally
$U(4N_{hf})$ gauge invariant under
transformations which only depend on the components
$x^{i}:=x^{\mu}e_{\mu}^{i}$.
 We shall also require the phenomenological
lagrangian
to be local gauge invariant under the corresponding transformations.

With the above modifications (i) and (ii) we shall apply the rules [18]
to
the case {\sl G}$=U(4N_{hf})$, {\sl H}$=U(2N_{hf})\otimes U(2N_{hf})$.
Let us first associate to the
currents (3.8) fields in the phenomenological lagrangian which have the
same transformation properties under {\sl H}
 $$ \eqalign{
 H^{ab} \rightarrow
\bar h^{a}i\gamma_5 p_{+} h^{b} \quad\,,\quad\quad &
 {H^{ab}}^{i} \rightarrow
\bar h^{a}i\el^{i} p_{+} h^{b} \,,\cr
 {H^{ba}}^{\ast} \rightarrow
\bar h^{b}i\gamma_5 p_{-} h^{a} \quad\,,\quad\quad &
 {{H^{ba}}^{i}}^{\ast} \rightarrow
-\bar h^{b}i\el^{i} p_{-} h^{a} \, . \cr } \eqno (4.1)
$$
We build up the following object
$$  {\rm H}= i\gamma_5
p_{-}H-i\el_{i}p_{-}H^{i}+i\gamma_5p_{+}H^{\dagger}
+i\el_{i}p_{+}{H^{i}}^{\dagger} \,,   \eqno (4.2)
$$
$${\rm \bar H}:=\gamma^0{\rm H}^{\dagger}\gamma^0 ={\rm H} \, ,$$
where we use matrix notation for $H^{ab}$ and ${H^{ab}}^{i}$. ${\rm H}$
 transforms
under the unbroken subgroup as follows
$$    {\rm H}\rightarrow h{\rm H}h^{-1} \quad , \quad h\in
U(2N_{hf})\otimes
U(2N_{hf}) \, .
\eqno (4.3) $$
We assign non-linear transformations under the full group $U(4N_{hf})$
in the standard manner [18]
$$g(\theta)e^{\rm H} =:e^{\rm H^{\prime}}h({\rm H},\theta)\,,$$
$$g\in U(4N_{hf}) \quad ,\quad h\in U(2N_{hf})\otimes U(2N_{hf})
\quad ,\quad e^{\rm H} \in
 U(4N_{hf}) / U(2N_{hf})\otimes U(2N_{hf}) \, ,
\eqno
(4.4) $$
where ${\rm H^{\prime}}$ is the transformed field. Then
$$e^{\rm H} \longrightarrow e^{\rm H^{\prime}}=ge^{\rm H}h^{-1} =h
e^{\rm H} \bar g \, ,\eqno (4.5)
 $$
where $\bar g =\gamma^0g^{\dagger}\gamma^0 $. The following property
holds

$$ \v e^{\rm H} =e^{\rm -H}\v \,,
\eqno (4.6)
$$
which implies that
$$ S:= e^{2{\rm H}}\v=\v e^{-2{\rm H}} \quad , \quad  S^2=1 \quad ,
\quad
   e^{2{\rm H}} \v \longrightarrow ge^{2{\rm H}}\v g^{-1}\, .
\eqno (4.7) $$

Because of the local gauge symmetry we can only build the following
connection and covariant tensor
$$   V:={1\over 2} \left( e^{-{\rm H}}v.\partial e^{\rm H} + e^{\rm
H}v.\partial
e^{-{\rm H}} \right) \quad , \quad V \longrightarrow
hVh^{-1}+hv.\partial
h^{-1} \quad , \quad \v V=V\v \,,
$$
$$   A:={1\over 2} \left( e^{-{\rm H}}v.\partial e^{\rm H} - e^{\rm
H}v.\partial e^{-{\rm H}} \right) \quad ,\quad A \longrightarrow
hAh^{-1} \quad\quad\quad , \quad \v A= -A\v \,,
 \eqno (4.8)
$$
$$ v.\partial S = e^{\rm H} A e^{\rm H} \v \, .$$

Notice that any derivative with respect to $x^{i}:=e^{i}_{\mu}x^{\mu}$
acting on functions of $x^{i}$ which are not scalars will not be
covariant under the local transformations.

The $u(4N_{hf})$ algebra and the HQET lagrangian are invariant
under the following discrete symmetry
$$   e_{\mu}^{i}\rightarrow -e_{\mu}^{i} \quad , \quad
    v_{\mu} \rightarrow -v_{\mu} \,,
 \eqno (4.9) $$
which is reminiscent of charge conjugation.
They are also invariant under the
$SO(3)$ transformations $e_{\mu}^{i}\rightarrow R^{i}_{j}e_{\mu}^{j}$
and, of course, under Lorentz transformations if we assign
$v_{\mu}\rightarrow {\Lambda_{\mu}}^{\nu}v_{\nu}$,
$e_{\mu}^{i} \rightarrow {\Lambda_{\mu}}^{\nu}e_{\nu}^{i}$.
 All these symmetries
  should
also be implemented in the effective lagrangian.

We can start at this point the construction of the effective
lagrangian, order by order in derivatives, using the objects defined
above. At first order it turns out that there is no invariant term.
Still there is a term which is invariant up to a total derivative. It
reads
 $$      Tr(\v V)\simeq
-4{\rm tr}(
H^{\dagger}v.\partial H   +
{H^{i}}^{\dagger}v.\partial H^{i}) + ... \;,
$$
$$ Tr(\v V) \longrightarrow Tr(\v V) +Tr(\v h v.\partial h^{-1}) \, .
 \eqno (4.10)
$$
$Tr$ means trace over flavour and Dirac indices whereas ${\rm tr}$
means trace over flavour indices only. We keep $tr$ for trace over
Dirac indices only. It
is not difficult to prove that $Tr(\v h v.\partial h^{-1})$ is indeed a
total derivative. This is analogous to the case of the Heisenberg
ferromagnet where the leading order term in the effective lagrangian
for the Goldstone mode is also invariant up to a total derivative [20].
Then at leading order the long distance properties of heavy quarkonia
are governed by a single constant. At next to leading order we
have the term
 $$                    Tr(AA)\simeq
-4{\rm tr}(
 v.\partial H^{\dagger}v.\partial H   +
v.\partial {H^{i}}^{\dagger}v.\partial H^{i}) + ...\; .
  \eqno (4.11)
$$
Terms containing $x^{i}$ derivatives start appearing at sixth order.
Notice that there is no vertex involving an odd number of fields.
This holds at any order in derivatives and it is a consequence of the
separate conservation of the number of heavy quarks and antiquarks.

For convenience we normalize the effective lagrangian as follows
$$ -i{f_{H}^2\over 4}Tr(\v V)= i{\rm tr}(\Pi^{\dagger}v.\partial \Pi +
{\Pi^{i}}^{\dagger} v.\partial \Pi^{i} ) + ...\;,$$
$$ H={\Pi\over f_{H}} \quad ,\quad
H^{i}={\Pi^{i} \over f_{H}} \, .
\eqno (4.12)
$$
$f_{H}^{2}$ is a dimension $3$ parameter of the order of
$\Lambda_{QCD}^3$.
The effective lagrangian built above makes sense by itself as a toy
model.
 If we ignore the matching with high energies we can
withdraw some consequences out of the lowest order lagrangian. These
and
the $1/m$ corrections to this toy model are worked out in the
appendix B.

\medskip
Let us next discuss how to represent quark currents in the effective
lagrangian. Consider
$$ j^{ab}_{\Gamma}=\bar h^{a}\Gamma h^{b} \, .
\eqno (4.13)$$
Let us introduce a source $a_{\Gamma}^{ab}$ for each of these currents
and write all possible currents up in the lagrangian
$$
L_v = i \bar h \v v _ {\mu} D ^ {\mu} h  + \bar h \v a h \,,$$
$$ a:= \sum_{\Gamma} a_{\Gamma}^{ab} \v \Gamma \, .
\eqno (4.14) $$
$L$ is now locally invariant under $U(4N_{hf})$ if we assign to $a$ the
transformation property
$$ a \longrightarrow g a g^{-1} + g iv.\partial g^{-1} \, .
\eqno (4.15) $$
At the hadronic level we may also require local gauge invariance upon
the introduction of $a$. This is easily achieved by changing
$v.\partial$ into $v.\partial -i a$ in the definition of $V$ in (4.8).
We obtain
$$ L=-i{f_{H}^2\over 4}\left[ Tr(\v V) -iTr(a S) \right] \, .
 \eqno (4.16) $$
Then we may identify
$$ \bar h^{a}\Gamma h^{b} \longrightarrow -{f_{H}^2\over 4}
Tr\left(\Gamma T^{ab}e^{2{\rm H}}\right) \, ,
\eqno (4.17)$$
where $T^{ab}$ is the zero matrix in flavor space except for a $1$  in
 row $a$ column $b$. It is interesting to observe that the $U(4N_{hf})$
symmetry is so large that any bilinear current of the kind (4.13) can be
written in terms of a generator of the $U(4N_{hf})$ symmetry.
This is the actual reason why the identification
(4.17) does not involve any extra unknown parameter. It is analogous to
the case of the vector and axial-vector currents in the Chiral
Lagrangian [21].
\medskip
Let us next calculate for further convenience the correlators (3.9) and
(3.10) in the hadronic effective lagrangian. For (3.9) we have
$$\eqalign{ G_{\Gamma \Gamma^{\prime}}(k)&=
\int d^4xe^{-ik.x}
\langle 0 \vert T\left\{
{{\bar h}^{a}_{-}}\Gamma
{h^{b}_{+}}(0)
{{\bar h}^{b}_{+}}  \Gamma^{\prime}
{h^{a}_{-}}(x)
\right\}\vert 0 \rangle          \cr
=&
\int d^4xe^{-ik.x}
\langle 0 \vert T\left\{
\Bigl[-{f_{H}^2\over 4}Tr(p_{-}\Gamma p_{+}T^{ab} e^{2{\rm
H}(0)})\Bigr]
\Bigl[-{f_{H}^2\over 4}Tr(p_{+}\Gamma^{\prime}p_{-} T^{ba} e^{2{\rm
H}(x)}) \Bigr]\right\}\vert 0 \rangle          \cr
\simeq     &
\int d^4xe^{-ik.x}
\langle 0 \vert T\left\{
\Bigl[-{f_{H}^2\over 4}Tr(p_{-}\Gamma p_{+}T^{ab} 2{\rm H}(0))\Bigr]
\Bigl[-{f_{H}^2\over 4}Tr(p_{+}\Gamma^{\prime}p_{-} T^{ba} 2{\rm
H}(x))\Bigr] \right\}\vert 0 \rangle          \cr
   = &-i{f_{H}^2 \over 2}tr\left(p_{+}\Gamma^{\prime}p_{-}\Gamma \right)
{1\over v.k+i\epsilon } \, .}
\eqno (4.18)
$$
For (3.10) we have
$$              \eqalign{
G_{\Gamma \Gamma^{\prime} \Gamma^{\prime\prime}} (\kp_1,\kp_2)=
&
\int d^4x_1
d^4x_2 e^{i\kp_1x_1-i\kp_2x_2}
\langle 0\vert T\left\{
{\bar h}^{a}_{-}\Gamma^{\prime\prime}  h^{b}_{+}(x_1)
{\bar h}^{b}_{+}\Gamma h^{c}_{+}(0)
{\bar h}^{c}_{+}\Gamma^{\prime} h^{a}_{-}(x_2)
\right\} \vert 0 \rangle
  \cr
= &
\int d^4x_1
d^4x_2 e^{i\kp_1x_1-i\kp_2x_2}
\langle 0\vert T\biggl\{
\Bigl[-{f_{H}^2\over 4}Tr(p_{-}\Gamma^{\prime\prime}p_{+} T^{ab}
e^{2{\rm H}(x_1)})\Bigr] \cr & \times
\Bigl[-{f_{H}^2\over 4}Tr(p_{+}\Gamma p_{+}T^{bc} e^{2{\rm H}(0)})\Bigr]
\Bigl[-{f_{H}^2\over 4}Tr(p_{+}\Gamma^{\prime}p_{-} T^{ca} e^{2{\rm
H}(x_2)})\Bigr] \biggr\} \vert 0 \rangle
  \cr
\simeq &
\int d^4x_1
d^4x_2 e^{i\kp_1x_1-i\kp_2x_2}
\langle 0\vert T\biggl\{
\Bigl[-{f_{H}^2\over 4}Tr(p_{-}\Gamma^{\prime\prime}p_{+} T^{ab} 2{\rm
H}(x_1))\Bigr] \cr &  \times
\Bigl[-{f_{H}^2\over 4}Tr(p_{+}\Gamma p_{+}T^{bc} 2{\rm H}^2(0))\Bigr]
\Bigl[-{f_{H}^2\over 4}Tr(p_{+}\Gamma^{\prime}p_{-} T^{ca} 2{\rm
H}(x_2))\Bigr] \biggr\} \vert 0 \rangle
  \cr
 = &
 {f_{H}^2 \over 2}
tr\left(p_{-} \Gamma^{\prime\prime} p_{+} \Gamma p_{+}
\Gamma^{\prime}\right)
{1 \over v.\kp_1
+i\epsilon} \,
{1 \over v.\kp_2
+i\epsilon} \, .}
\eqno (4.19)
$$
Notice at this point that we may obtain (3.9) and (3.10) from (4.18) and
(4.19) by taking $
f_{H}^2/2\rightarrow
N_{c} \mu^3 / 6\pi^2 $. Hence $f_{H}^2$ at the hadronic level
plays
the role of the cut-off $\mu$ at quark level. Observe also that the
dependence on the $\Gamma$-matrices in (4.18)-(4.19) is explicit.
All decay constants and matrix elements of
bilinear currents are given in terms of the only non-perturbative
parameter $f_{H}$. This is a direct consequence of
the $U(4N_{hf})$
symmetry being spontaneously broken down to $U(2N_{hf})\otimes
U(2N_{hf})$.

\bigskip
{\bf 5. Example: the decay constant, $\bf f_{\Upsilon}$}
\medskip

{\bf 5.1. Separating and evaluating off-shell and on-shell
contributions}
\medskip

Consider the current-current correlator $$  G_{\Gamma}(p):=
\int d^4x
e^{ipx}\langle 0\vert T\left\{
\bar Q^{a} \Gamma Q^{b}(0)
\bar Q^{b}\bar\Gamma Q^{a}(x)
\right\} \vert 0 \rangle \,,\eqno (5.1)
$$
$$p=-m_{ab,n}v-k \quad ,\quad k\rightarrow 0 \,. $$
We separate $$ \bar Q^{a} \Gamma Q^{b}=
(\bar Q^{a} \Gamma Q^{b})_{on}+
(\bar Q^{a} \Gamma Q^{b})_{off} \,.\eqno (5.2)
$$
Where $(\bar Q^{a} \Gamma Q^{b})_{on}$ and  $(\bar Q^{a} \Gamma
Q^{b})_{off}$ means that both heavy quark fields in the
current have momenta almost on-shell and off-shell respectively. Our
goal is to obtain a representation in terms of the HQET of any Green
function containing an $(\bar Q^{a} \Gamma Q^{b})_{on}$. In order
to enforce "on-shellness" it is convenient to make the substitution
$$
\int d^4x  (\bar Q^{b}\bar\Gamma Q^{a}(x))_{on} e^{ipx}
\longrightarrow
\int d^4x_1{\bar Q}^{b i_1}_{\alpha_1}(x_1) e^{ip_1x_1}
\int d^4x_2Q^{a i_2}_{\alpha_2}(x_2) e^{ip_2x_2}
(\bar \Gamma)_{\alpha_1\alpha_2}  \delta_{i_1 i_2}\,,
                           \eqno (5.3)
$$
$$
\eqalign{
& p_1=-(m_{a}-{m_{a}\over m_{ab}}E_{ab,n})v-k^{\prime}_1 \,,
 \cr & p_2=-(m_{b}-{m_{b}\over m_{ab}}E_{ab,n})v-k^{\prime}_2 \,,
\cr & k=k_1^{\prime}+k_2^{\prime}\,,
\cr & k_1^{\prime} , k_2^{\prime} \rightarrow 0 }
\eqno (5.4)
$$
and see whether the new Green function admits a representation in terms
of the HQET.
This is nothing but
the calculations carried out above. Then we undo (5.3) by putting the
fields depending on $x_1$ and $x_2$ in the HQET at the same point $x$.
We have (from (2.1), (2.13) and (3.11))
 $$ \eqalign{  & \int d^4x
e^{ipx}\langle 0\vert T\left\{
(\bar Q^{a} \Gamma Q^{b}(0))_{off}
(\bar Q^{b}\bar\Gamma Q^{a}(x))_{on}
\right\} \vert 0 \rangle \cr & =
\int d^4x e^{-ikx}\langle
0\vert T\left\{ C_{\Gamma}\bar
h^{a}_{-}{\Gamma}{h}^{b}_{+}(0)\bar {h}^{b}_{+}\bar \Gamma
{h}^{a}_{-}(x)\right\} \vert 0 \rangle\,.}
\eqno (5.5)
$$
Analogously, using (2.14), (2.16) and (3.13) we have
$$
\eqalign{  &
\int d^4x
e^{ipx}\langle 0\vert T\left\{
(\bar Q^{a} \Gamma Q^{b}(0))_{on}
(\bar Q^{b}\bar\Gamma Q^{a}(x))_{on}
\right\} \vert 0 \rangle \cr & =
\int d^4x
 e^{-ikx}
\langle 0 \vert T\Biggl\{
{\bar h^{a}_{-}}\Gamma
{h^{b}_{+}}(0)
{\bar h^{b}_{+}}\bar \Gamma
{h^{a}_{-}}(x) \cr &
\times i\int d^4y \left( -{1\over 2 N_{c}}
\tilde \Psi_{ab,n}^{\ast}(0)
\tilde \Psi_{ab,n}(0) \right)
\sum_{\Gamma_{n}=i\gamma_5p_{-} , i\el^{i}p_{-}}
\bar h^{b}\Gamma_{n} h^{a}(y)
iv.\partial
(\bar h^{a}\bar \Gamma_{n} h^{b}(y)) \Biggr\}
 \vert 0 \rangle \,.}
\eqno (5.6)
$$
The contribution involving only off shell quarks has the familiar form
$$
\eqalign{ & \int d^4x
e^{ipx}\langle 0\vert T\left\{
(\bar Q^{a} \Gamma Q^{b}(0))_{off}
(\bar Q^{b}\bar\Gamma Q^{a}(x))_{off}
\right\} \vert 0 \rangle \cr & =
-iN_{c}
 tr( \Gamma p_{+}\bar \Gamma p_{-})
\vert \Psi_{ab,n}(0)\vert^2
{1\over v.k +i\epsilon}\,.}
\eqno (5.7)
$$
The expressions (5.5) and (5.6) correspond to corrections
$O(\Lambda^{3}_{QCD}a^{3}_{ab,n})$
 and
$O(\Lambda^{6}_{QCD}a^{6}_{ab,n})$
respectively to the leading result (5.7), $a_{ab,n}\sim n/(\alpha
\mu_{ab})$ is the Bohr radius. Since we are only
interested
in the leading non-perturbative corrections we shall neglect (5.6) in
the following. Let us only remark that the hadronization of the
four quark operator in (5.6) introduces new parameters. This is because
it is not a generator of the $U(4N_{hf})$ symmetry as the currents of
the kind (4.17) are.

The r.h.s. of (5.5) can be hadronized and calculated using the
effective lagrangian discussed in section 4. From (4.18) we obtain
$$
\eqalign{  & \int d^4x
e^{ipx}
\langle 0\vert T\left\{
(\bar Q^{a} \Gamma Q^{b}(0))_{off}
(\bar Q^{b}\bar\Gamma Q^{a}(x))_{on}
\right\} \vert 0 \rangle \cr & =
{-i\over 2}
 tr( p_{-}\Gamma p_{+}\bar \Gamma )
\tilde \Psi_{ab,n}^{\ast}(0)\Psi_{ab,n}(0) f_{H}^2
{1\over v.k +i\epsilon}\,.}
\eqno (5.8)
$$
Notice that the result is spin independent and the flavor dependence
resides only in the wave function, which is known.
We finally obtain
$$  \vert f_{\psi_{Q}(n)}\vert^2=4m_{ab,n}
\left(N_{c}
\vert \Psi_{ab,n}(0)\vert^2+{1\over 2}
\left(
\tilde \Psi_{ab,n}^{\ast}(0)\Psi_{ab,n}(0) +
 \Psi_{ab,n}^{\ast}(0)\tilde \Psi_{ab,n}(0)\right)
f_{H}^2\right) \,,
\eqno (5.9)
$$
$$ \vert f_{\eta_{Q}(n)}\vert ={\vert f_{\psi_{Q}(n)}\vert \over
m_{ab,n}} \,.$$
Notice that the non-perturbative correction we find to the decay
constant is
$O(\Lambda_{QCD}^3a_{ab,n}^3)$
and hence presumably more important that the correction arising from the
multipole expansion which is
$O((\Lambda_{QCD}a_{ab,n})^4/\a^2)$ [8,9] (we count the quark
condensate as $O(\Lambda_{QCD}^4)$).

\medskip
{\bf 5.2 Cut-off independence}
\medskip

Let us next discuss the important issue of the cut-off independence.
Even
though we have not written it down explicitely, the introduction of a
cut-off to separete almost on-shell momenta from off-shell momenta is
necessary. Of course, the final results must not depend on the
particular value of the cut-off.
At the short distance end of the calculation, the cut-off must exclude
momenta which are almost on-shell. This is easily achieved by cutting
off small momenta from the wave function
 $$  \Psi_{ab,n} (0)=
 \int {d^3\vec k \over (2\pi)^3} \tilde \Psi_{ab,n} (\vec k)
\longrightarrow
 \int_{\mu} {d^3\vec k \over (2\pi)^3} \tilde \Psi_{ab,n} (\vec k)
=:\Psi_{ab,n}^{(\mu)}(0)\,,
\eqno (5.10)
$$
where $\mu$ is a symmetric IR cut-off in three momentum.
The wave functions in (5.9) must be understood as the cut-off wave
functions (5.10). On the HQET side the cut-off must be ultraviolet. It
has already been displayed in
the leading order perturbative calculation at quark level
in section 3. In particular, from (3.9) we obtain
$$
\eqalign{  & \int d^4x
e^{ipx}\langle 0\vert T\left\{
(\bar Q^{a} \Gamma Q^{b}(0))_{off}
(\bar Q^{b}\bar\Gamma Q^{a}(x))_{on}
\right\} \vert 0 \rangle \cr & =
{-i\over 2}
N_{c}tr(p_{-} \Gamma p_{+}\bar \Gamma)
\tilde \Psi_{ab,n}^{\ast}(0)\Psi_{ab,n}(0)
 ({ \mu^3 \over 6\pi^2})
{1 \over v.k
+i\epsilon}\,.}
\eqno (5.11)
$$
\medskip
This strong cut-off dependence, however, is
 totally compensated by (5.10). Indeed, once (5.10) is used
we have
$$
\eqalign{      {d\over d\mu}
\vert \Psi_{ab,n}^{(\mu)}(0)\vert^2= & -{\mu^2 \over 2\pi^2}
\left(
\tilde \Psi_{ab,n}^{\ast} (\mu) \Psi_{ab,n}^{(\mu)}(0)
+{ \Psi_{ab,n}^{(\mu)}}^{\ast} (0) \tilde \Psi_{ab,n}(\mu) \right) \cr
= & - {\mu^2\over 2\pi^2}
\biggl(
\tilde \Psi_{ab,n}^{\ast} (0) \Psi_{ab,n}(0)
+ \Psi_{ab,n}^{\ast} (0) \tilde \Psi_{ab,n}(0) \cr &+ O\left((\mu
a_{ab,n})^2\right) \biggr)\,,
\cr & \cr
{d\over d\mu}
\left(
\tilde \Psi_{ab,n}^{\ast}(0)\Psi_{ab,n}^{(\mu)}(0)
 ({ \mu^3 \over 6\pi^2})
\right)
             =                    &
\tilde \Psi_{ab,n}^{\ast}(0)\Psi_{ab,n}(0)
 { \mu^2 \over 2\pi^2}\left[1
+ O\left((\mu a_{ab,n})^2\right)   \right]\,,
\cr & \cr
{d\over d\mu}
\left(
{\Psi_{ab,n}^{(\mu)}}^{\ast}(0)\tilde \Psi_{ab,n}(0)
 ({ \mu^3 \over 6\pi^2})
\right)
             =                    &
\Psi_{ab,n}^{\ast}(0) \tilde \Psi_{ab,n}(0)
 { \mu^2 \over 2\pi^2}\left[1
+ O\left((\mu a_{ab,n})^2\right)    \right]\,.}
\eqno (5.12)
$$
\medskip
Notice that the way in which the cut-off dependence cancels is
remarkable. The strong cut-off dependence of (5.11) was first found in
[11]. It was not clear at all which short distance contribution it
should
cancel against. (5.10) gives the solution to that puzzle. It is
apparent from  (5.8) and (5.11)
that $f_{H}$ in the hadronic theory plays the role of the UV cut-off
in the HQET at quark level. From
(5.12) it is clear that the cut-off $\mu$ must be much smaller than the
invers Bohr
radius. Therefore our formalism becomes exact in the following
situation
$$ m_{a} , m_{b} \gg
1/ a_{ab,n}
\gg \mu \gg
\Lambda_{QCD} \gg \kp \,,
 $$
$$
{\mu_{ab}\alpha^2(1/ a_{ab,n}) \over n^2}
\gg  \kp \,.
   \eqno (5.13)
$$
Furthermore, we have to assume that $\mu$ can be taken large enough so
that we may
enter the asymptotic freedom regime from the HQET side. Otherwise the
matching we have carried out at tree level would not make much
sense.

From the discussion above it should also be clear that (5.9) can be
written
in a cut-off independent way at $O((\mu a_{ab,n})^3)$ by just replacing
$$ f_{H}^2 \longrightarrow {\bar f_{H}^2 } := {f_{H}^2 }
-{N_{c}\mu^3\over 3\pi^2} \,,\eqno (5.14)
$$
where ${\bar f_{H}}^2 $ need not be positive.

\medskip
{\bf 5.3. Physical state normalization}
\medskip

There is still a subtle point which makes eq. (5.9) with the
replacement (5.14) not quite correct. It has to do with the
normalization of physical states. It will be clear later on (see eq.
(6.14) below) that the states we obtain by this procedure do not have
the standard relativistic normalization as they are supposed to.
When we evaluate the Green function (5.1) we
insert resolutions of the identity which are approximated by
Coulomb-type states. This is all right. However the low momentum tale
of these
states is cut-off and substituted by a quantity evaluated using the
effective hadronic theory. After doing so there is no guarantee that the
resolution of the identity we introduced is still properly normalized.
This can be fixed up by changing
$$
 \sum_{n}\int{d^3 \vec P_{n} \over (2\pi )^3 2P_{n}^0}\vert n \rangle
\langle n \vert \longrightarrow
 \sum_{n}\int{d^3 \vec P_{n} \over (2\pi )^3 2P_{n}^0}
\vert n \rangle
\langle n \vert^{(\mu)}
N_{n}(\mu ,f_{H})
\eqno (5.15)
$$
where $
\vert n \rangle
\langle n \vert^{(\mu)}
$ symbolises the cut-off Coulomb states whose low energy tale is
evaluated in the hadronic effective theory. We present a heuristic
calculation of $
N_{n}(\mu ,f_{H})$.

 \medskip
We start from
the Coulomb-type bound
state (2.5) and
separate high and low relative momentum according to
$$
\vert \Gamma_{n} ,
\vec P_{n}=m_{ab,n}\vec v
\rangle
=
\vert \Gamma_{n} ,
\vec P_{n}=m_{ab,n}\vec v
\rangle^{k>\mu}
+
\vert \Gamma_{n} ,
\vec P_{n}=m_{ab,n}\vec v
\rangle^{k<\mu}
\eqno (5.16)
$$
The high momentum part of the physical state can be well approximated by
the Coulomb-type
contribution so we may leave it as it stands. However, the low momentum
part receives non-perturbative corrections, which we evaluate using the
effective hadronic lagrangian.

We proceed as follows. Since
$a_{ab,n}\mu\ll 1$, we can approximate the low momentum region by
$$
\eqalign{
\vert \Gamma_{n} ,
\vec P_{n}=m_{ab,n}\vec v
\rangle^{k<\mu}
\simeq &
{v^0\over \sqrt{N_{c}} }
 {\tilde \Psi_{ab,n}(\vec 0)\over \sqrt{2m_{a}v^02m_{b}v^0}}
{m_{ab}^{3\over 2}\over m_{ab,n}} \cr & \times \int^{k<\mu} {d^3\vec k
\over (2\pi)^3}
\sum_{\alpha ,\beta} \bar u^{\alpha} (p_1) \Gamma_{n} v^{\b}(p_2)
a^{\dagger}_{\a}(p_1)b^{\dagger}_{\b}(p_2) \vert 0\rangle\,,}
\eqno (5.17)
 $$
Observe now that (5.17) is nothing but the integral of a local HQET
current.  $$
\vert \Gamma_{n} ,
\vec P_{n}=m_{ab,n}\vec v
\rangle^{k<\mu}   \simeq
{v^0\over \sqrt{N_{c}} }
  \tilde \Psi_{ab,n}(\vec 0)
{m_{ab}^{3\over 2}\over m_{ab,n}} \int d^3\vec x
e^{-ikx}
\bar h^{a}\Gamma_{n} h^{b}(x) \vert 0 \rangle \,,
\eqno (5.18)
 $$
where $k \to 0$ and only low momenta are allowed.

At this point, we can hadronize the current (see (4.17)) and calculate
the low momentum contribution to $N_{n}(\mu ,f_{H})$
$$
\eqalign{ & ^{k<\mu}\langle \Gamma_{n} ,
\vec P_{n}=m_{ab,n}\vec v
\vert
 \Gamma_{n} ,
\vec P_{n}^{\prime}=m_{ab,n}\vec v^{\prime}
\rangle^{k<\mu}   \cr &              =
2m_{ab,n}v^0 (2\pi)^3\delta^{(3)}(m_{ab,n}(\vec v -\vec v^{\prime}))
{f^2_{H}\over 2N_{c}} \vert \tilde \Psi_{ab,n}(\vec 0) \vert^2 \,.}
\eqno (5.19)
$$
Then, putting together high and low momentum contributions, we have
$$
\eqalign{ & \langle \Gamma_{n} ,
\vec P_{n}=m_{ab,n}\vec v
\vert
 \Gamma_{n} ,
\vec P_{n}^{\prime}=m_{ab,n}\vec v^{\prime}
\rangle   \cr &              =
2m_{ab,n}v^0 (2\pi)^3\delta^{(3)}(m_{ab,n}(\vec v -\vec v^{\prime}))
\left[
\int_{k>\mu} {d^3\vec k
\over (2\pi)^3} \vert \tilde \Psi_{ab,n}(\vec k) \vert^2
+{f^2_{H}\over 2N_{c}} \vert \tilde \Psi_{ab,n}(\vec 0) \vert^2\right]
\cr &  =
2m_{ab,n}v^0 (2\pi)^3\delta^{(3)}(m_{ab,n}(\vec v -\vec v^{\prime}))
\left[1
+{\bar f^2_{H}\over 2N_{c}} \vert \tilde \Psi_{ab,n}(\vec 0)
\vert^2\right] \,.} \eqno (5.20)
$$
Where $\bar f^2_{H}$ is defined in (5.14) and
notice that the result is cut-off independent.

Finally, the normalization factor reads

$$N_{n}(\mu ,f_{H})=
{1\over{1+
{\bar f^2_{H}\over 2N_{c}} \vert \tilde \Psi_{ab,n}(\vec 0) \vert^2}}
\,.
\eqno (5.21)
$$

$N_{n}(\mu ,f_{H})     $
can also be obtained from requiring
$$
      \langle  \Gamma_{n} ,
\vec P_{n}=m_{ab,n}\vec v
\vert
 \int d^3\vec x \bar Q^{b}\gamma^0 Q^{b}(\vec x)
\vert \Gamma_{n} ,
\vec P_{n}^{\prime}=m_{ab,n}\vec v^{\prime}
\rangle                   =
2m_{ab,n}v^0 (2\pi)^3\delta^{(3)}(m_{ab,n}(\vec v -\vec v^{\prime}))
\eqno (5.22)
$$
as we shall see later on. Once we have taken into account the correct
normalization (5.9) reads
$$  \eqalign{
  \vert f_{\psi_{Q}(n)}\vert^2= & 4m_{ab,n}
\Biggl[N_{c}
\vert \Psi_{ab,n}(0)\vert^2
+{1\over 2}
\left(
\tilde \Psi_{ab,n}^{\ast}(0)\Psi_{ab,n}(0) +
 \Psi_{ab,n}^{\ast}(0)\tilde \Psi_{ab,n}(0)\right)
{\bar f_{H}}^2
 \cr &
-
\vert \Psi_{ab,n}(0)\vert^2
\vert \tilde \Psi_{ab,n}(0)\vert^2
{{\bar f_{H}}^2 \over 2} \Biggr] }
\eqno (5.23)
$$

We shall relegate to section 7 the discussion on the applicability of
the limit (5.13) and formula (5.23) to physical situations.

  \bigskip
 {\bf 6. Matrix elements at zero recoil}
 \bigskip

We are interested in Green functions of the kind
$$
G_{\Gamma \Gamma^{\prime} \Gamma^{\prime\prime}} (p_1,p_2) =
\int d^4x_1
d^4x_2 e^{ip_1x_1+ip_2x_2}
\langle 0\vert T\left\{
\bar Q^{a} \Gamma^{\prime\prime} Q^{b}(x_1)
\bar Q^{b}\Gamma Q^{c}(0)
\bar Q^{c}\Gamma^{\prime} Q^{a}(x_2)
\right\} \vert 0 \rangle \,.
\eqno (6.1)
$$
For the momentum range
$$ p_1=m_{ab,n}v+k_1^{\prime}
\quad , \quad
 p_2=-m_{ac,m}v-k_2^{\prime}\,.
$$
$$ \kp_1\, ,\, \kp_2 \longrightarrow 0
\eqno (6.2)
$$
We separate each current in almost on-shell momenta and off-shell
momenta according to (5.2). The leading contribution is given by the
term
$$
\eqalign{ &
\,G_{\Gamma \Gamma^{\prime} \Gamma^{\prime\prime}} (p_1,p_2)
\cr &= \int d^4x_1
d^4x_2 e^{ip_1x_1+ip_2x_2}
\langle 0\vert T\left\{
(\bar Q^{a} \Gamma^{\prime\prime} Q^{b}(x_1))_{off}
(\bar Q^{b}\Gamma Q^{c}(0))_{off}
(\bar Q^{c}\Gamma^{\prime} Q^{a}(x_2))_{off}
\right\} \vert 0 \rangle \cr &
=N_{c} tr(p_{-} \Gamma^{\prime\prime} p_{+} \Gamma p_{+}
\Gamma^{\prime})
\Psi_{ac,m}^{\ast}(0) \Psi_{ab,n}(0)
\int {d^3\vec k\over (2\pi)^3}
\tilde \Psi_{ab,n}^{\ast}(\vec k)\tilde \Psi_{ac,m}(\vec k)
{1 \over v.\kp_1
+i\epsilon} \,
{1 \over v.\kp_2
+i\epsilon} }
\eqno (6.3)
$$
and the next to leading contribution by the term
$$
G_{\Gamma \Gamma^{\prime} \Gamma^{\prime\prime}}^{on} (p_1,p_2):=
G_{\Gamma \Gamma^{\prime} \Gamma^{\prime\prime}}^{on,1} (p_1,p_2)+
G_{\Gamma \Gamma^{\prime} \Gamma^{\prime\prime}}^{on,2} (p_1,p_2)+
G_{\Gamma \Gamma^{\prime} \Gamma^{\prime\prime}}^{on,3} (p_1,p_2)\,,
\eqno (6.4)
$$
$$
\eqalign{ & \,G_{\Gamma \Gamma^{\prime}
\Gamma^{\prime\prime}}^{on,1} (p_1,p_2) \cr & = \int d^4x_1
d^4x_2 e^{ip_1x_1+ip_2x_2}
\langle 0\vert T\left\{
(\bar Q^{a} \Gamma^{\prime\prime} Q^{b}(x_1))_{on}
(\bar Q^{b}\Gamma Q^{c}(0))_{off}
(\bar Q^{c}\Gamma^{\prime} Q^{a}(x_2))_{off}
\right\} \vert 0 \rangle\,,}
\eqno (6.5)
$$
$$
\eqalign{ & \,G_{\Gamma \Gamma^{\prime}
\Gamma^{\prime\prime}}^{on,2} (p_1,p_2) \cr & = \int d^4x_1
d^4x_2 e^{ip_1x_1+ip_2x_2}
\langle 0\vert T\left\{
(\bar Q^{a} \Gamma^{\prime\prime} Q^{b}(x_1))_{off}
(\bar Q^{b}\Gamma Q^{c}(0))_{on}
(\bar Q^{c}\Gamma^{\prime} Q^{a}(x_2))_{off}
\right\} \vert 0 \rangle\,,}
\eqno (6.6)
$$
$$
\eqalign{ & \,G_{\Gamma \Gamma^{\prime}
\Gamma^{\prime\prime}}^{on,3} (p_1,p_2) \cr & =\int d^4x_1
d^4x_2 e^{ip_1x_1+ip_2x_2}
\langle 0\vert T\left\{
(\bar Q^{a} \Gamma^{\prime\prime} Q^{b}(x_1))_{off}
(\bar Q^{b}\Gamma Q^{c}(0))_{off}
(\bar Q^{c}\Gamma^{\prime} Q^{a}(x_2))_{on}
\right\} \vert 0 \rangle\,.}
\eqno (6.7)
$$
The calculation of (6.5) and (6.7) is analogous to the ones carried out
in section 2. We obtain
$$        \eqalign{
G_{\Gamma \Gamma^{\prime} \Gamma^{\prime\prime}}^{on,1} (p_1,p_2)=
&
\int d^4x_1
d^4x_2 e^{i\kp_1x_1-i\kp_2x_2}
iC_1 \langle 0\vert T\left\{
\bar h^{a}_{-}\Gamma^{\prime\prime}h^{b}_{+}(x_1) \bar h^{b}_{+}\Gamma
p_{+}\Gamma^{\prime} h^{a}_{-}(0) \right\} \vert 0 \rangle
\cr & \times \int d^4q{e^{iqx_2}\over v.q+i\epsilon} \,,    }
 \eqno (6.8)$$
$$ C_1=
\Psi_{ac,m}^{\ast}(0) \tilde \Psi_{ab,n}(0)
\int {d^3\vec k\over (2\pi)^3}
\tilde \Psi_{ab,n}^{\ast}(\vec k)\tilde \Psi_{ac,m}(\vec k) \,,
$$
$$ \eqalign{
G_{\Gamma \Gamma^{\prime} \Gamma^{\prime\prime}}^{on,3} (p_1,p_2)=
&
\int d^4x_1
d^4x_2 e^{i\kp_1x_1-i\kp_2x_2}
iC_3 \langle 0\vert T\left\{
\bar h^{a}_{-}\Gamma^{\prime\prime} p_{+}
\Gamma h^{c}_{+}(0)
\bar h^{c}_{+}\Gamma^{\prime} h^{a}_{-}(x_2)
\right\} \vert 0 \rangle
\cr & \times \int d^4q{e^{-iqx_1}\over v.q+i\epsilon}\,,  }
\eqno (6.9)$$
$$ C_3=
\tilde \Psi_{ac,m}^{\ast}(0)  \Psi_{ab,n}(0)
\int {d^3\vec k\over (2\pi)^3}
\tilde \Psi_{ab,n}^{\ast}(\vec k)\tilde \Psi_{ac,m}(\vec k) \,.
$$
Notice that (6.5) and (6.7) can not be written in terms of local Green
functions in the HQET. One propagator must be kept explicit.

The calculation  of (6.6) is more subtle. We describe it in some detail
in the Appendix B. We obtain
$$
G_{\Gamma \Gamma^{\prime} \Gamma^{\prime\prime}}^{on,2} (p_1,p_2)=
\int d^4x_1
d^4x_2 e^{i\kp_1x_1-i\kp_2x_2}
C_2 \langle 0\vert T \left\{
\bar h^{a}_{-}\Gamma^{\prime\prime}
h^{b}_{+}(x_1)
\bar h^{b}_{+}\Gamma h^{c}_{+}(0)
\bar h^{c}_{+}\Gamma^{\prime} h^{a}_{-}(x_2)
\right\} \vert 0 \rangle\,,
\eqno (6.10)$$
$$ C_2=
\Psi_{ac,m}^{\ast}(0)  \Psi_{ab,n}(0)
\tilde \Psi_{ac,m}^{\ast}(0)  \tilde \Psi_{ab,n}(0) \,.
$$
This term is the only one in (6.4) which remains in the matrix elements
(see (6.14) below).

 We calculate (6.8)-(6.10) using the hadronic
effective lagrangian (see formulas  (4.18) and (4.19)). We obtain
$$
G_{\Gamma \Gamma^{\prime} \Gamma^{\prime\prime}}^{on,1} (p_1,p_2)=
C_1{f_{H}^2\over 2}
tr(p_{-} \Gamma^{\prime\prime} p_{+} \Gamma p_{+}
\Gamma^{\prime})
{1 \over v.\kp_1
+i\epsilon} \,
{1 \over v.\kp_2
+i\epsilon} \,,
\eqno (6.11)
$$
$$
G_{\Gamma \Gamma^{\prime} \Gamma^{\prime\prime}}^{on,2} (p_1,p_2)=
C_2
{f_{H}^2\over 2}
tr(p_{-} \Gamma^{\prime\prime} p_{+} \Gamma p_{+}
\Gamma^{\prime})
{1 \over v.\kp_1
+i\epsilon} \,
{1 \over v.\kp_2
+i\epsilon} \,,
\eqno (6.12)
$$
$$
G_{\Gamma \Gamma^{\prime} \Gamma^{\prime\prime}}^{on,3} (p_1,p_2)=
C_3{f_{H}^2\over 2}
tr(p_{-} \Gamma^{\prime\prime} p_{+} \Gamma p_{+}
\Gamma^{\prime})
{1 \over v.\kp_1
+i\epsilon} \,
{1 \over v.\kp_2
+i\epsilon} \,.
\eqno (6.13)
$$
The matrix element at zero recoil then reads
$$
\eqalign{ &
      \langle  \Gamma_{n} ,
\vec P_{n}=m_{ab,n}\vec v
\vert
\bar Q^{b}\Gamma Q^{c}(0)
\vert \Gamma_{m} ,
\vec P_{m}=m_{ac,n}\vec v
\rangle                   \cr & =-
\sqrt{m_{ab,n}m_{ac,m}} tr( \bar \Gamma_{n} \Gamma
\Gamma_{m})
\left(
\int {d^3\vec k\over (2\pi)^3}
\tilde \Psi_{ab,n}^{\ast}(\vec k)\tilde \Psi_{ac,m}(\vec k)
      +
{f_{H}^2\over 2N_{c}}
\tilde \Psi_{ac,m}(0)  \tilde \Psi_{ab,n}^{\ast}(0)
                                  \right)
\,,}
\eqno (6.14)
$$
$\Gamma_{n}=i\gamma_5p_{-} , i\el^{i}p_{-}$ for the pseudoscalar and
vector particle respectively.
The integral in (6.14) must be understood with an infrared cut-off
$\mu$.
From (6.14) it is apparent that our physical states are not properly
normalized. Indeed, for $b=c$ and $\Gamma=\gamma^0$ one should obtain
(5.22) but one does not. The reason for this has been discussed at the
end of sect. 5. The solution consist of introducing the
normalization factor $N_{n}(\mu , f_{H})$ defined in (5.21).
The properly normalized result reads
$$
\eqalign{ &
      \langle  \Gamma_{n} ,
\vec P_{n}=m_{ab,n}\vec v
\vert
\bar Q^{b}\Gamma Q^{c}(0)
\vert \Gamma_{m} ,
\vec P_{m}=m_{ac,m}\vec v
\rangle                    =-
\sqrt{m_{ab,n}m_{ac,m}} tr( \bar \Gamma_{n} \Gamma
\Gamma_{m})   \cr & \times
\Biggl[
\int {d^3\vec k\over (2\pi)^3}
\tilde \Psi_{ab,n}^{\ast}(\vec k)\tilde \Psi_{ac,m}(\vec k)
 \left( 1-
{{\bar f_{H}}^2\over 4N_{c}}
\vert \tilde \Psi_{ab,n}(0) \vert^{2}
-
{{\bar f_{H}}^2\over 4N_{c}}
\vert \tilde \Psi_{ac,m}(0) \vert^{2} \right) \cr &
     +
{{\bar f_{H}}^2\over 2N_{c}}
\tilde \Psi_{ac,m}(0)  \tilde \Psi_{ab,n}^{\ast}(0)
                                  \Biggr]
\,,}
\eqno (6.15)
$$

 Notice that the non-perturbative correction depends only on
a
single parameter ${\bar f}^2_{H}$ which may be extracted from the decay
constants
calculated in section 5. This is a non-trivial prediction which turns
out to be a direct consequence of the
 $U(4N_{hf})$
symmetry being spontaneously broken down to $U(2N_{hf})\otimes
U(2N_{hf})$.

    \bigskip\bigskip

{\bf 7. Applications}
    \bigskip
If the charm and bottom mass were large enough we could
apply the results above to the physics of
$\Upsilon$, $\eta_{b}$, $B_{c}$, $ B_{c}^{\ast}$, $J/\Psi$ and
$\eta_{c}$. (The top is believed to be too heavy to form hadronic
bound states and will be ignored). We analyse in this section whether
this is so or not.
In the systems where the formalism
actually applies,
 we are mainly interested in estimating the importance
of
the new non-perturbative contribution
 rather than in obtaining accurate results. The latter
is a much harder task which is definitely beyond the scope of the
present work.

 \medskip

Let us first focus on bottom. The fact that the almost 'on-shell'
momentum excitations in
heavy quarkonium are Goldstone modes [11] implies that the $\Upsilon$
and $\eta_{b}$ spectrum does not received additional non-perturbative
contributions. We may then extract the bottom mass from the $\Upsilon$
mass
by means of the formulas given in [8,9] which take into account the
leading
order in the multipole expansion. Since we have established a link
between quarkonium and the HQET we can next use $m_{b}$ to extract $\bar
\Lambda $, the non-perturbative parameter relating the mass of the
$B$-meson to $m_{b}$. Moreover, taking into account that $\bar \Lambda$
is flavour independent, we may next extract the charm mass $m_{c}$. We
summarize the results in the Table I.
\medskip
In Table I the values we obtain for $m_{b}$
are about a $3\%$ lower than those obtained in QCD sum rules [22] but
compatible
with a recent QCD-based evaluation [23] and with the lattice
calculation [24]. The values we obtain
for $\bar \Lambda$  are a bit lower but otherwise
compatible with those
extracted from QCD sum rules [6]. Our values for $m_{c}$ are again
about a $6\%$ lower than the typical values in QCD sum rules [22].
 We should emphasize that our
numbers in Table I are model independent.

\medskip

 We can next extract the non-perturbative parameter
 ${\bar f_{H}}^2$ from
 $f_{\Upsilon}$ (this is done in Table II).
We use the following formula
$$\eqalign{ & {f_{\Upsilon}} =2\sqrt{3m_{\Upsilon}}\Psi_{bb,0}(0) \cr &
\times \left[ 1+{\tilde \Psi_{bb,0}(0)\bar f_{H}^2\over6\Psi_{bb,0}(0)}
-{\vert\tilde \Psi_{bb,0}(0)\vert^{2}\bar f_{H}^2\over12}
-{8\a(m_{b})\over3\pi}+8.77m_{b}^2<B^2>({a_{bb,0}\over2})^6 \right]
\,,}\eqno (7.1)
$$
where the 1-loop QCD corrections and the leading correction from the
multipole expansion [9] \footnote{*}{We use the formula given in
ref. [9] which differs from the ones in ref. [8].} are taken into
account.

The numbers in Table II are very sensitive to the scale at
which
$\a $ is taken. Notice that we choose $\a=\a (1/a_{bb,0})$ in the Bohr
radius and binding energy but $\a=\a (m_{b})$ in the 1-loop
perturbative correction included in (7.1).
From Table II we see that for the actual values of $m_{b}$ and
$\Lambda_{QCD}=100\, ,\, 150 MeV$
the 'on-shell' contribution ($\bar f_{H}$) does not dominate
over the condensate, but it is certainly sizeable.
For $\Lambda_{QCD}=200 MeV$ all corrections are about the same order
and for any value of $m$ the 'on-shell' contribution dominates over the
condensate.

\medskip

Observe that the conditions (5.13), in
particular $a_{bb,0}^{-1}\gg \mu \gg
\Lambda_{QCD}$, may be considered as reasonable well fulfilled if we
take the cut-off $\mu \sim 700$ MeV (see table III below).

\medskip

Let us next turn our attention to charm. The charm mass is known not to
be heavy
enough as for the multipole expansion to work [8]. This means that the
non-perturbative contribution overwhelms the perturbative one. Therefore
any approximation whose leading order is a perturbative contribution,
like our approach, will not be able to say much about charmonium. In
particular, for the 'on-shell' contributions the difficulty lies on the
second last
condition in (5.13) being fulfilled. There is little room to accomodate
the cut-off $\mu$ between the invers Bohr radius and $\Lambda_{QCD}$ as
should be clear from Table III.
We refrain ourselves from giving any numbers for charmonium.

 Unfortunately, the situation is
not much better
for the $B_{c}$, which has received considerable attention lately
[25-27].
Nonetheless,
once we have $\bar f_{H}^2$,
 we shall give some numbers in this case in Table IV.

From Table IV we see that for
$\Lambda_{QCD}=100\, ,\, 150 MeV$ the contribution of the condensate is
too large for the approach to be reliable. For
$\Lambda_{QCD}=200 MeV$
 we are at the boundary of its validity
 since
the 'on-shell' correction is large.
We may thus
give a rought estimate for $f_{B_{c}}$  only for $\Lambda_{QCD}\sim 200
MeV $, which
turns out to be compatible with the estimate
obtained by
QCD sum rules [26], but about a $30\%$ lower than
potential model estimates [27].

\medskip
From the Table V it follows that the new non-perturbative
contribution is not very important in
the matrix elements between $\Upsilon-B_{c}$ states.

The decay constants and matrix elements above receive
contributions from corrections of several types:
 \medskip
(i) QCD perturbative corrections to the Coulomb potential
$O(\alpha (1/a_{n}))$.
 These have been evaluated at one loop level in
[28] (see also [23]).
\medskip

(ii) Relativistic corrections to the Coulomb potential
$O(\alpha (1/a_{n}))$ (see also [28,23]).
\medskip

(iii) QCD perturbative corrections to the Green functions
$O(\alpha (m))$.
These corrections have been taken into account in (7.1).
 They
correspond to the only QCD corrections in heavy-light systems. In
our case they are important for the calculation of
 matrix elements at non-zero recoil.
\medskip

(iv) Non-perturbative corrections arising from the multipole expansion
in the off-shell momentum region
$O(\Lambda_{QCD}^4a_{n}^4/\alpha^2 (1/a_{n}))$ [8,9].
These corrections have also been taken into account in (7.1).
\medskip

(v) Finite mass corrections
$O(\Lambda_{QCD}^2/m)$
in the hadronic HQET lagrangian.
\medskip

    \bigskip\bigskip

 {\bf 8. Conclusions}
         \bigskip
We have demostrated that, contrary to the common belief, HQET
techniques are also useful for the study of systems composed of two
heavy quarks. In particular, we have identified new
nonperturbative contributions to the
decay constants and to certain matrix elements which
are
described by a hadronic lagrangian based on the HQET.
All these new contributions
are parametrized at leading order by a single constant $f_{H}$.
 This is non-trivial and can be traced back to the fact that a
 $U(4N_{hf})$
symmetry is spontaneously broken down to $U(2N_{hf})\otimes
U(2N_{hf})$.
\medskip

It is remarkable that strong cut-off dependences coming from
a totally different origin match perfectly. Indeed, at the off-shell end
the cut-off arises from an integral over a Coulomb type
wave function, whereas at the on-shell end it arises from a Feynman
integral.
        \medskip

We should also stress that we have been able to put in the same context
(i.e. the HQET) both heavy-heavy and heavy-light systems. This allows
for a model independent determination of heavy quark masses from
quarkonium, which may then be used to extract the parameter $\bar
\Lambda$
relating the mass of the heavy-light systems to the mass of the heavy
quark.

\medskip

 As far as practical applications is concerned, our formalism is
suitable for the ground state of the $\Upsilon$ and $\eta_{b}$ family.
Unfortunately the charm mass is too small for the formalism to become
applicable in general to $J/\Psi$ and $B_{c}$ systems. Nevertheless one
may stretch it in some cases
to obtain information on the mass and decay constant of the latter.

\medskip

We have presented a technique which allows to disentangle the on-shell
contributions from the rest and match them to the HQET. The matching
has
been carried out at tree level. We have already shown in [29] that the
matching also goes through at one loop level. Nevertheless, a word of
caution is needed. It would be desirable to have a more direct and
systematic derivation of these results from QCD. Progress in this
direction is being made [30].

\medskip
 Let us finally mention that
the hadronic HQET lagrangian can easily incorporate heavy-light mesons.
The formalism can then be extended to the calculation of matrix
elements between quarkonium and heavy-light systems. The leading
non-perturbative contributions to those are also given by $f_{H}$ and
another non-perturbative parameter which is related to heavy-light decay
constants.
Non-recoil contributions can also be evaluated within the formalism.
 \bigskip \bigskip
{\bf Acknowledgements}
\bigskip
This work has been supported in part by CICYT grant AEN93-0695. A.P.
acknowledges a fellowship from CIRIT. J.S. has benefited from the
Fermilab Summer Visitors Program. He thanks E. Eichten for illuminating
conversations and J.M. Sanchis for sending his papers in [7] prior
to publication. Thanks are also given to J.L. Goity and P. Pascual
for the critical reading of the manuscript.
  \bigskip
 {\bf Appendix A: a toy model}
\bigskip

Because of the similarity, both in physics and techniques, to the
Chiral
perturbation theory it is interesting to consider a toy model which
contains the on-shell contributions only. At quark level the model is
described by the HQET with quarks and antiquarks with the same velocity
of sect. 3. At hadronic level is described by the effective hadronic
lagrangian of sect. 4.

Within this model, the
interactions between
 ($\eta_{Q},\eta_{Q^{\prime}}$),
 ($\eta_{Q},\psi_{Q^{\prime}}$),
 ($\psi_{Q},\psi_{Q^{\prime}}$),
when the two particles move roughtly at the same velocity, are described
by
a single unknown constant. This is analogous to the fact that at lowest
order in 3-flavour chiral perturbation theory the elastic scattering of
($\pi , \pi $), ($K , K$) and ($\pi , K$) is also described by a
single constant. When heavy-light mesons are included in the
effective
lagrangian the same constant describes the elastic scattering of
heavy-light mesons with quarkonium. This is also analogous to the
fact that the local vertex $\pi$-$\pi$-$N$-$N$ at leading order in the
Chiral Lagrangian is
described by the same constant as the ($\pi , \pi $) elastic scattering.
Let us mention at this point that when one actually calculates the
scattering
amplitudes, one obtains zero. This has to do with the universality of
the
leading order effective lagrangians for Goldstone modes [18,19,20]. Any
theory undergoing a $U(4N_{hf})$ spontaneous symmetry breaking down to
$U(2N_{hf})\otimes
U(2N_{hf})$ has the same low energy effective lagrangian (4.12) provided
the
rest of the symmetries in the theory are also the same. It was shown in
[11], that
even when the gluons are switched off the spontaneous symmetry breaking
occurs in the HQET. In that case there is no interaction in the
fundamental theory and hence it is not surprising that the scattering
amplitudes in the effective lagrangian vanish. Universality
implies
that there will be vanishing scattering amplitudes when the gluons are
switched on as well.

\medskip
Within this model one can also treat $1/m$ corrections in a similar way
that small quarks masses are dealt with in Chiral perturbation theory.
At the quark level
the leading $1/m$ corrections to the HQET are given by a kinetic term
$$ -\sum_{a=1}^{N_{hf}} {1\over 2m_{a}}D_{i}\bar h^{a} D_{i}
h^{a} \eqno (A.1) $$
and a spin breaking term
$$ \sum_{a=1}^{N_{hf}} {1\over 4m_{a}} \bar h^{a} S^{l} G^{l}
h^{a}
$$ $$  G^{l}=-{1\over 2} \epsilon^{jkl}e^{\mu}_{j}e^{\nu}_{k} G_{\mu\nu}
\, . \eqno (A.2)
$$
The kinetic term (A.1) does not break the global $U(2N_{hf})\otimes
U(2N_{hf})$ symmetry but
it breaks its local version. In order to construct at the hadronic level
terms which break the $U(4N_{hf})$ symmetry in the same fashion as (A.1)
does, we introduce the $u(4N_{hf})$-valued sources $\phi$ and $a_{i}$
transforming as
$$\phi \longrightarrow  g\phi g^{-1} \,,$$
$$a_{i} \longrightarrow g a_{i} g^{-1}+ g\partial_{i}g^{-1}\,.
\eqno (A.3)$$
Then the term
$$       - d_{i}\bar h \v \phi d_{i} h \,,$$
$$  d_{i}h:=( D_{i} +a_{i})h \quad , \quad
  d_{i}\bar h\v:= D_{i}\bar h\v - \bar h\v a_{i}  \eqno
(A.4) $$
is on one hand invariant under $U(4N_{hf})$ and on the other reduces to
(A.1) upon setting
$$ a_{i}=0 \quad , \quad \phi=
\pmatrix{{1\over 2m_{a}} & &  \cr
 & {1\over 2m_{b}} & \cr & & \ddots \cr}
 \v \,. \eqno (A.5)
$$
At the hadronic level, we must then construct invariant terms linear in
$\phi$, which may also contain $a_{i}$. Up to two space derivatives we
have
 $$\eqalignno{ & tr(S\phi) \,,& (A.6) \cr & tr(S\phi
d_{i}Sd_{i}S)\,,&(A.7) \cr & tr(S\phi) tr(d_{i}Sd_{i}S)\,,& (A.8) }
$$
$$ d_{i}S :=\partial_{i} S+ a_{i}S-Sa_{i} \,.$$
We have not written down terms which coincide or vanish upon using
(A.5).

For the spin breaking term (A.2) we may introduce a
$u(4N_{hf})$-valued source $R^{l}$ transforming as
$$ R^{l}\longrightarrow gR^{l}g^{-1}
\eqno (A.9)$$
so that (A.2) is substituted by
$$
   \bar h\v R^{l} G^{l} h \,.
\eqno (A.10) $$
We recover (A.2) upon setting
$$ R^{l}=
\pmatrix{{1\over 4m_{a}} & &  \cr
 & {1\over 4m_{b}} & \cr & & \ddots \cr}
\v S^{l} \,.
\eqno (A.11)
$$
There are no terms at the hadronic level with
the same symmetry transformation properties at lower orders in
derivatives. The first possible term appears at third order.

Therefore the leading $1/m$ corrections introduce $3$ new
parameters. (A.6) provides a mass term $O(\Lambda_{QCD}^2/m)$ and
(A.7)-(A.8)
give rise to the usual non-relativistic kinetic term.
The procedure above can easily be
extended to any finite order in $1/m$.

 \bigskip
{\bf Appendix B}
\bigskip

We present in this appendix  some technical details on the evaluation of
the off-shell short distance effects carried out in sect. 6.

Consider the following matrix element
$$
\langle  \Gamma_{n} ,
m_{ab,n}\vec v
\vert
\bar Q^{b}\Gamma Q^{c}(x)
\vert \Gamma_{m} ,
m_{ac,n}\vec v
\rangle\,.
\eqno (B.1)
$$
Since two different bound states are involved, it is not clear {\it a
priori} which CM dependence one should substract before using (2.5).
Nevertheless, translation invariance implies that the result of
the calculation must fulfill
 $$        \eqalign{
\langle  \Gamma_{n} ,
m_{ab,n}\vec v
\vert
\bar Q^{b}\Gamma Q^{c}(x+a)
\vert \Gamma_{m} ,
m_{ac,n}\vec v
\rangle
=                   &
e^{im_{ab,n}v.a-im_{ac,m}v.a}  \cr &
\times \langle  \Gamma_{n} ,
m_{ab,n}\vec v
\vert
\bar Q^{b}\Gamma Q^{c}(x)
\vert \Gamma_{m} ,
m_{ac,n}\vec v
\rangle   \,.                  }
\eqno (B.2)
$$
We also have
$$ \eqalign{
\langle  \Gamma_{n} ,
m_{ab,n}\vec v
\vert
\bar Q^{b}\Gamma Q^{c}(x)
\vert \Gamma_{m} ,
m_{ac,n}\vec v
\rangle
=  &
e^{im_{ab,n}v.\xi-im_{ac,m}v.\xi} \cr &
\times \langle  \Gamma_{n} ,
m_{ab,n}\vec v
\vert
\bar Q^{b}\Gamma Q^{c}(x-\xi)
\vert \Gamma_{m} ,
m_{ac,n}\vec v
\rangle \,. }
\eqno (B.3)
$$
If we assign $\xi \rightarrow \xi + a$ under translations (B.3)
fulfils
(B.2). If we also require $\xi$ to be a linear function of $x$,
then necessarily $\xi=x$ and the result is well defined.
$$  \eqalign{ &
\langle  \Gamma_{n} ,
m_{ab,n}\vec v
\vert
\bar Q^{b}\Gamma Q^{c}(x)
\vert \Gamma_{m} ,
m_{ac,n}\vec v
\rangle \cr &
=
e^{im_{ab,n}v.x-im_{ac,m}v.x}
\langle  \Gamma_{n} ,
m_{ab,n}\vec v
\vert
\bar Q^{b}\Gamma Q^{c}(0)
\vert \Gamma_{m} ,
m_{ac,n}\vec v
\rangle
 \cr & =
e^{im_{ab,n}v.x-im_{ac,m}v.x}\left(-
 tr( \bar \Gamma_{n} \Gamma
\Gamma_{m})
\int {d^3\vec k\over (2\pi)^3}
\tilde \Psi_{ab,n}^{\ast}(\vec k)\tilde \Psi_{ac,m}(\vec k)
\right)\,.
}
\eqno (B.4)
$$
Consider next
$$
\langle  \Gamma_{n} ,
m_{ab,n}\vec v
\vert
{\bar Q}^{b i_3}_{\a_3}(x_3) Q^{c i_4}_{\a_4}(x_4)
\vert \Gamma_{m} ,
m_{ac,n}\vec v
\rangle \,.
\eqno (B.5)
$$
We are in a similar situation as above. However now translation
invariance does not fix completely the result. Under the same
assumptions we obtain
$$  \eqalign{ &
\langle  \Gamma_{n} ,
m_{ab,n}\vec v
\vert
{\bar Q}^{b i_3}_{\a_3}(x_3) Q^{c i_4}_{\a_4}(x_4)
\vert \Gamma_{m} ,
m_{ac,m}\vec v
\rangle
\cr & = e^{i(\a x_3+(1-\a)x_4
)(E_{ac,m}-E_{ab,n})+im_{b}v.x_3-im_{c}v.x_4}
\Bigl(-{1\over N_{c}}
 (\Gamma_{m} \bar \Gamma_{n}
)_{\a_4 \a_3}
\delta_{i_3 i_4}
\cr &
\times \int {d^3\vec k\over (2\pi)^3}
\tilde \Psi_{ab,n}^{\ast}(\vec k)\tilde \Psi_{ac,m}(\vec k)
e^{i(\vec k .\vec v (x^0_3-x^0_4) - (\vec x_3-\vec x_4) (\vec k +{\vec k
.\vec v \over 1+v^0}\vec v)) }\Bigr) \,,}
\eqno (B.6)
$$
where $\alpha$ is arbitrary and parametrizes the ambiguity. Usually one
never runs into calculations of the kind (B.5) but rather of matrix
elements of currents as in (B.1), which are not ambiguous. We find
expresions like (B.5) in our calculation because we insist in enforcing
"on-shellness" in certain currents. In our concrete case we have a
current with a momentum insertion
$$
\bar Q^{b}\Gamma Q^{c}(x)e^{ip.x} \quad\quad, \quad
p=(-m_{b}+m_{c}+E_{ab,n}-E_{ac,m})v \,.
\eqno (B.7)
 $$
In order to enforce on-shellness we substitute it by
$$
{\bar Q}^{b i_3}_{\a_3}(x_3) Q^{c i_4}_{\a_4}(x_4)
e^{ip_3 x_3 +ip_4 x_4}
(\Gamma)_{\a_3 \a_4}
\delta_{i_3 i_4} \,,
$$
$$ p_3=-(m_{b}-{m_{b}\over m_{ab}}E_{ab,n})v-\kp_3 \quad , \quad
 p_4=(m_{c}-{m_{c}\over m_{ac}}E_{ac,m})v+\kp_4
\quad , \quad \kp_3 ,\kp_4 \rightarrow 0 \,,
\eqno (B.8)
 $$
as mentioned in (5.3). However in doing so there is a momentum
missmatch $
({m_{a}\over m_{ab}}E_{ab,n} -  {m_{a}\over
m_{ac}}E_{ac,m})v
$ which should be fixed somehow in order to get back (B.7) in the
$x_3=x_4=x$
limit. The most general way of distributing this momentum missmatch
between $x_3$ and $x_4$ is by inserting in (B.8)
$$
 e^{i(\b x_3 +(1-\b )x_4)
({m_{a}\over m_{ab}}E_{ab,n} -  {m_{a}\over
m_{ac}}E_{ac,m})v
}\,.
\eqno (B.9)
$$
Any $\beta$ is equaly good since we are eventually interested in the
limit $x_3=x_4=x$. Notice that the ambiguity in $\alpha$ in (B.6) is
proportional to the ambiguity in $\beta$ in (B.9). Since we can choose
$\beta$ at will, we do it in such a way that the dependence in both
$\alpha$ and $\beta$ cancels. This is how we are able to obtain a
representation of (6.6) in terms of the HQET (6.10).

\bigskip
\centerline{\bf References}
\bigskip

\item{[1]} M.B. Voloshin and M.A. Shifman, {\it Yad. Fiz.} {\bf 45}
(1987) 463 [Sov. J. {\it Nucl. Phys.} {\bf 45} (1987) 292].
 \item{} H.D.
Politzer and M.B. Wise, {\it Phys. Lett.} {\bf B206}
(1988) 681;  {\it Phys. Lett.} {\bf B208} (1988) 504.

\item{[2]} N. Isgur and M.B. Wise, {\it Phys. Lett.} {\bf B232}
(1989) 113;  {\it Phys. Lett.} {\bf B237} (1990) 527.

\item{[3]} E. Eichen and B. Hill, {\it Phys. Lett.} {\bf B234} (1990)
511.

\item{[4]} H. Georgi,
{\it Phys. Lett.} {\bf B240} (1990) 447.

\item{[5]} B. Grinstein, {\it Nucl. Phys.} {\bf B339} (1990) 253.

\item{[6]} B. Grinstein, in {\it Proceedings of the Workshop on High
Energy Phenomenology}, Mexico City Mexico, Jul. 1-12, 1991, 161-216
and in {\it Proceedings Intersections between particle and nuclear physics}
Tucson 1991, 112-126.
\item{} T. Mannel, {\it Chinese Journal of Physics} {\bf 31} (1993) 1.
\item{} M. Neubert, {\it 'Phys. Rep.'} {\bf245} (1994) 259.

\item{[7]} R. Casalbuoni, A. Deandrea, N. Di Bartolomeo, R. Gatto, F.
Ferruglio and G. Nardulli,
{\it Phys. Lett.} {\bf B309} (1993) 163;
{\it Phys. Lett.} {\bf B302} (1993) 95.
\item{} M.A. Sanchis,
{\it Phys. Lett.} {\bf B312} (1993) 333; {\it Z. Phys. C} {\bf 62}
(1994) 271.

\item{[8]} H. Leutwyler,
{\it Phys. Lett.} {\bf B98} (1981) 447.

\item{[9]} M. B. Voloshin,
{\it Nucl. Phys. } {\bf B154} (1979) 365; Sov. J. Nucl. Phys., {\bf36},
143 (1982).

\item{[10]} J. Soto and R. Tzani, {\it Phys. Lett.} {\bf B297} (1992)
358.

\item{[11]} J. Soto and R. Tzani,  {\it Int. J. Mod. Phys.} {\bf A9}
(1994) 4949.

\item{[12]} A. Das and V.S. Mathur,
{\it Phys. Rev.} {\bf D49} (1994) 2508.
\item{} A. Das and M. Hott, {\it 'Chiral invariance of massive
fermions'}, UR-1352 preprint.

 \item{[13]} T. Mannel, W. Roberts and Z.
Ryzak, {\it Nucl. Phys. } {\bf B368} (1992) 204.

\item{[14]}
A. F. Falk, H. Georgi, B. Grinstein and M.B. Wise, {\it Nucl. Phys.}
{\bf B343} (1990) 1.

\item{[15]} B. Grinstein, W. Kilian, T. Mannel and M. Wise,
{\it Nucl. Phys.} {\bf B363} (1991) 19.
\item{} W. Kilian, T. Mannel and T. Ohl,
{\it Phys. Lett.} {\bf B304} (1993) 311.

\item{[16]} W. Lucha, F.F. Schorbel and D. Gromes, {\it Phys. Rep.}{\bf
200} (1991) 127.

 \item{[17]} F.E. Close and Z. Li,
{\it Phys. Lett.} {\bf B289} (1992) 143.

\item{[18]} S. Coleman, J. Wess and B. Zumino, {\it Phys. Rev. } {\bf
D177} (1969) 2239.
\item{} C. Callan, S. Coleman, J. Wess and B. Zumino,
{\it Phys. Rev.}
{\bf D177} (1969) 2247.

\item{[19]} H. Leutwyler, {\it Ann. Phys.} {\bf 235} (1994) 165.

\item{[20]} H. Leutwyler, {\it Phys. Rev.} {\bf D49} (1994) 3033.

\item{[21]} J. Gasser and H. Leutwyler, {\it Ann. Phys. } {\bf(N.Y.)158}
(1984) 142;
{\it Nucl. Phys.}
{\bf B250} (1985) 465.

\item{[22]} S. Narison, {\it Lecture Notes in Physics}, {\bf Vol.
26},{\it 'QCD Spectral Sum Rules'} (World Scientific, Singapore, 1989).

\item{[23]} S. Titard and F.J. Yndur\'ain,
{\it Phys. Rev.}
{\bf D49} (1994) 6007.

\item{[24]} C.T.H. Davis, K. Hornbostel, A. Langau, G.P. Lepage, A.
Liedsey, C.J. Morningstar, J. Shigenitsu and J. Sloan, {\it Phys. Rev.
Lett.} {\bf 73} (1994) 2654.

\item{[25]} E. Jenkins, M. Luke, A. V. Manohar and M. Savage,
{\it Nucl. Phys.}
{\bf B390} (1993) 463.

\item{[26]} E. Bagan, H.G. Dosch, P. Gosdzinsky, S. Narison and J.-M.
Richard, {\it Z. Phys.} {\bf C64} (1994) 57.

\item{[27]} E.J. Eichten and C. Quigg, {\it Phys. Rev.} {\bf D49}
(1994) 5845.
\item{} V.V. Kiselev, A.K. Likhoded and A.V. Tkabladze, {\it '$B_{c}$
spectroscopy'}, IHEP 94-51 preprint.

\item{[28]} J. Pantaleone, S.-H. Henry Tye and Y. J. Ng ,
{\it Phys. Rev.}
{\bf D33} (1986) 777.

\item{[29]} A. Pineda and J. Soto, {\it Phys. Lett.} {\bf B361} (1995)
95.

\item{[30]} A. Pineda and J. Soto, in progress.

\eject
\footline={\hss\folio\hss}
\baselineskip=20pt

Table I. We use
$\Lambda_{QCD}$ as an input and take the one loop running coupling
constant $\a$ at the scale
of the invers Bohr radius, i.e. $\a =\a (1/a_{bb,0})$. For the gluon
condensate we take the fix value $<B^2>=(585 MeV )^4$.
The error in $m_{b}$ has been taken from estimations of the hiperfine
splittings $O(\a^2)$, which are also the main source of error in
$\bar \Lambda $. For $m_{c}$ the error comes both from $\bar \Lambda $
and the $1/m_{c}$ corrections. The last column gives our model
independent determination of $m_{B_{c}}$.
\bigskip\bigskip

Table II. We display the relative
weight, with its sign, of the 1-loop ($\a (m_{b})$), the condensate
($<B^2>$) and the
'on-shell' ($\bar f_{H}$) contribution with respect to the Coulomb type
contribution (normalized
to 1). The last columns display the mass $m_{cr}$ from which the
'on-shell' contribution dominates over the condensate and the value of
$({\bar f}^2_{H})^{1/3}$.
\bigskip\bigskip

Table III. We give the ${\bar c}c, {\bar b}c, {\bar b}b$ invers Bohr
radius as a function of $\Lambda_{QCD}$.
\bigskip\bigskip

Table IV. We display the analogous to
Table II for $B_{c}$. We have also given our predictions for
$f_{B_{c}}$ in the last column.
\bigskip\bigskip

Table V. We give the relative weight, with its sign, of the
'on-shell' contribution with respect to the Coulomb-type contribution
(normalized to $1$) in the matrix elements (6.15) between $\Upsilon
-B_{c}$ states.

\eject
\footline={\hss\folio\hss}
\baselineskip=20pt

$$\vbox{\tabskip=0pt \offinterlineskip
\halign to380pt{\strut#& \vrule#\tabskip=1em plus2em& \hfil#& \vrule#&
\hfil#& \vrule#& \hfil#& \vrule#& \hfil#& \vrule#&\hfil#&
\vrule#\tabskip=0pt\cr \noalign{\smallskip}
& \multispan7Table I.\hfil\cr \noalign{\hrule}
& & \omit\hidewidth $\Lambda_{QCD}$(MeV)\hidewidth& & \omit\hidewidth
$m_{b}$(MeV)\hidewidth& & \omit\hidewidth $\bar\Lambda$(MeV)\hidewidth&
& \omit\hidewidth $m_{c}$(MeV)\hidewidth& & \omit\hidewidth
$m_{B_c}$(MeV)\hidewidth& \cr \noalign{\hrule}
& & 200& & 4877$\pm$35& & 436$\pm$35& & 1539$\pm$70& &
6212$\pm$110& \cr \noalign{\hrule} & & 150& & 4843$\pm$35& &
470$\pm$35& & 1505$\pm$70& & 6242$\pm$110& \cr
\noalign{\hrule} & & 100& & 4802$\pm$35& & 511$\pm$35& &
1464$\pm$70& & 6312$\pm$110&
\cr\noalign{\hrule}
\noalign{\smallskip}}}$$

$$\vbox{\tabskip=0pt \offinterlineskip
\halign to391pt{\strut#& \vrule#\tabskip=1em plus2em& \hfil#& \vrule#&
\hfil#& \vrule#& \hfil#& \vrule#& \hfil#& \vrule#& \hfil#& \vrule#&
\hfil#& \vrule#\tabskip=0pt\cr \noalign{\smallskip}
& \multispan7Table II.\hfil\cr \noalign{\hrule}
& & \omit\hidewidth $\Lambda_{QCD}$(MeV)\hidewidth& & \omit\hidewidth
$\a (m_{b})$\hidewidth& & \omit\hidewidth $<B^2>$
\hidewidth& & \omit\hidewidth $\bar f_{H}$\hidewidth& &
\omit\hidewidth $m_{cr}$(GeV)\hidewidth& & \omit\hidewidth $({\bar
f}^2_{H})^{1/3}$(MeV)\hidewidth& \cr \noalign{\hrule} & & 200& & -0.19&
& 0.10&
& -0.11& & ---& & 260& \cr \noalign{\hrule} & & 150& & -0.17& & 0.19&
& -0.08& & 90& & 210& \cr
\noalign{\hrule} & & 100& & -0.15& & 0.41& & -0.12& &
160& & 210& \cr\noalign{\hrule}\noalign{\smallskip}}}$$

$$\vbox{\tabskip=0pt \offinterlineskip
\halign to300pt{\strut#& \vrule#\tabskip=1em plus2em& \hfil#& \vrule#&
\hfil#\hfil& \vrule#& \hfil#& \vrule#&
\hfil#& \vrule#\tabskip=0pt\cr \noalign{\smallskip}
& \multispan7Table III.\hfil\cr \noalign{\hrule}
& & \omit\hidewidth $\Lambda_{QCD}$(MeV)\hidewidth& & \omit\hidewidth
$1/a_{cc,0}$(MeV)\hidewidth& &
\omit\hidewidth $1/a_{bc,0}$(MeV)\hidewidth& & \omit\hidewidth
$1/a_{bb,0}$(MeV)\hidewidth& \cr \noalign{\hrule} & & 200& & 630& & 790&
& 1240& \cr \noalign{\hrule} & & 150& & 540& & 700& & 1120& \cr
\noalign{\hrule} & & 100& & 450& & 590& & 980& \cr
\noalign{\hrule}\noalign{\smallskip}}}$$

$$\vbox{\tabskip=0pt \offinterlineskip
\halign to300pt{\strut#& \vrule#\tabskip=1em plus2em& \hfil#& \vrule#&
\hfil#\hfil& \vrule#& \hfil#& \vrule#&\hfil#& \vrule#&
\hfil#& \vrule#\tabskip=0pt\cr \noalign{\smallskip}
& \multispan7Table IV.\hfil\cr \noalign{\hrule}
& & \omit\hidewidth $\Lambda_{QCD}$(MeV)\hidewidth& & \omit\hidewidth
$\a (2\mu_{bc})$\hidewidth& & \omit\hidewidth $<B^2>$
\hidewidth& & \omit\hidewidth $\bar f_{H}$\hidewidth& &
\omit\hidewidth $f_{B_c}$(MeV)\hidewidth& \cr \noalign{\hrule}
& & 200& & -0.24& & 0.35& & -0.44& & 370&\cr
\noalign{\hrule} & & 150& & -0.22& & 0.74& & -0.34& &
540& \cr
\noalign{\hrule} & & 100& & -0.19& & 1.93& & -0.54& &
780& \cr \noalign{\hrule} \noalign{\smallskip}}}$$

$$\vbox{\tabskip=0pt \offinterlineskip
\halign to300pt{\strut#& \vrule#\tabskip=1em plus2em& \hfil#& \vrule#&
\hfil#\hfil& \vrule#& \hfil#& \vrule#&\hfil#& \vrule#\tabskip=0pt\cr
\noalign{\smallskip}
& \multispan7Table V.\hfil\cr \noalign{\hrule}
& & \omit\hidewidth $\Lambda_{QCD}$(MeV)\hidewidth& & \omit\hidewidth
200\hidewidth& & \omit\hidewidth 150\hidewidth& &
\omit\hidewidth 100\hidewidth& \cr \noalign{\hrule}
& & $B_{c}-\Upsilon$& &
-0.10& & -0.08& & -0.14& \cr \noalign{\hrule} \noalign{\smallskip}}}$$
 \bigskip \bigskip

\end